\documentclass{WileyMSP-template}

\usepackage{amsfonts}
\usepackage{amssymb}
\usepackage{mathtools}

\usepackage{color}
\definecolor{purple}{rgb}{0.8,0,0.6}
\definecolor{PURPLE}{rgb}{0.8,0,0.6}
\definecolor{orange}{rgb}{1,0.55,0}
\definecolor{darkblue}{rgb}{0,0,0.55}
\definecolor{darkred}{rgb}{0.75,0,0}
\definecolor{gray}{rgb}{0.1,0.1,0.1}

\newcommand{\beqn}{\begin{eqnarray}}
\newcommand{\eeqn}{\end{eqnarray}}
\newcommand{\eq}[1]{(\ref{#1})}

\newcommand{\cL}{{\cal L}}

\newcommand{\vort}{{\mathrm{vort}}}

\newcommand{\latt}{{\mathrm{latt}\,}}

\newcommand{\Z}{{\mathbb Z}}
\newcommand{\bs}{\boldsymbol}

\newcommand{\avr}[1]{{\left\langle #1 \right\rangle}}

\begin{document}

\pagestyle{fancy}
\rhead{\includegraphics[width=2.5cm]{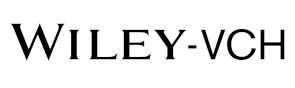}}

\title{Generation of electric current by magnetic field at the boundary: \\ quantum scale anomaly vs. semiclassical Meissner current outside of the conformal limit}

\maketitle


\author{Maxim Chernodub*}
\author{Vladimir Goy}
\author{Alexander Molochkov}


\dedication{} 

\begin{affiliations}
Prof. M. N. Chernodub\\
Institut Denis Poisson UMR 7013, Universit\'e de Tours, 37200 France\\
Email Address: maxim.chernodub@univ-tours.fr

Dr. V. A. Goy, Prof. A. V. Molochkov\\
Pacific Quantum Center, Far Eastern Federal University, 690950 Vladivostok, Russia

\end{affiliations}


\keywords{scalar QED, scale anomaly, magnetic field, superconducting, boundaries, lattice simulations.}

\begin{abstract}
The scale (conformal) anomaly can generate an electric current near the boundary of a system in the presence of a static magnetic field. The magnitude of this magnetization current, produced at zero temperature and in the absence of matter, is proportional to a beta function associated with the renormalization of the electric charge. Using first-principle lattice simulations, we investigate how the breaking of the scale symmetry affects this ``scale magnetic effect''  near a Dirichlet boundary in scalar QED (Abelian Higgs model). We demonstrate the interplay of the generated current with vortex excitations both in symmetric (normal) and broken (superconducting) phases and compare the results with the anomalous current produced in the conformal, scale-invariant regime. Possible experimental signatures of the effect in Dirac semimetals are discussed.
\end{abstract}

\section{Introduction}

Quantum anomalies lead to distinctive transport phenomena, such as the chiral magnetic effect and its generalizations, which produce various currents of electric and axial charges in diverse physical environments, including systems at finite density and finite temperature, at strong electromagnetic fields, and in vortical backgrounds~\cite{Kharzeev:2012ph,Kharzeev:2013ffa}.

The appearance of anomalous currents is directly related to breaking a classical symmetry by quantum fluctuations, with the most vivid examples given by the axial anomaly and the mixed axial-gravitational anomaly~\cite{Landsteiner:2011cp,Landsteiner:2012kd}. These phenomena involve the axial (chiral) degrees of freedom that are expected to apply to various physical environments, from quark-gluon plasma created in ultrarelativistic heavy-ion collisions~\cite{Huang:2015oca} to condensed matter environments~\cite{Landsteiner:2016led}.

The anomalous breaking of the classical conformal (scale) invariance~\cite{Capper:1973mv,Capper:1974ed,Deser:1976yx} can also be responsible for anomalous transport effects~\cite{Chernodub:2021nff}. For example, the scale electromagnetic effects generate an electric current and produce electric charge accumulation in electromagnetic backgrounds in curved spacetime~\cite{Chernodub:2016lbo}. In condensed matter context, the curved spacetime can be modeled as a temperature gradient that drives a system out of equilibrium~\cite{Luttinger:1964zz}. The anomalous phenomena are then seen as the Nernst effect~\cite{Chernodub:2017jcp} in which the magnitude of the anomalous electric current is controlled by the beta function related to the renormalization of the electric charge. A purely gravitational conformal anomaly can reveal itself in various solid-state systems by affecting, for example, thermodynamic characteristics such as pressure gradients~\cite{Chernodub:2019tsx} and thermal transport~\cite{Bermond:2022mjo,Northe:2022tjr}. 

In a bounded classically conformal system, the scale anomaly can generate another type of anomalous electric current which reveal itself in the presence of the background magnetic fields~\cite{Chu:2018ksb,Chu:2018ntx} (see also Ref.~\cite{McAvity:1990we} for an earlier study). The anomalous current flows along the boundary with a magnitude proportional to the beta function of the electric charge, similar to the currents generated by the scale electromagnetic effects in bulk. In addition, the conformal anomaly leads to the electric charge accumulation, which can have an observable imprint in Dirac semimetals at charge neutrality point~\cite{Chernodub:2019blw}. Anomaly-induced boundary phenomena also include a large variety of other effects associated with different symmetries or curved boundaries~\cite{Herzog:2017xha,Andrei:2018die,Chu:2020mwx,Hu:2020puq,Miao:2022oas}.

The appearance of electric boundary currents in the presence of the background magnetic field is in no way a new phenomenon. For example, the Meissner current can also be generated by the motion of the Cooper pairs in a thin layer of a superconductor in response to the applied magnetic field~\cite{Tinkham2004}. As the magnetic field penetrates the superconductor, it induces a circulating current in the material that flows in such a way as to generate an opposing magnetic field. This current is carried by the Cooper pairs, which can flow freely without any resistance due to their superconducting nature.

There are several conceptual differences between anomalous and superconducting boundary currents. The anomalous current is generated, as the name says for it, by the quantum anomaly in the vacuum. On the other hand, the Meissner current has a semiclassical non-anomalous nature and is produced in matter (in the presence of a superconducting condensate). Our paper investigates quantitative differences between these currents on the unified footing within the same system. 

We numerically study, from first principles, the generation of an electric boundary current in the lattice formulation of the (3+1) dimensional Abelian Higgs model (AHM) following the analysis of Ref.~\cite{Chernodub_2018ihb}. Our choice of this model is governed by its distinct features. First of all, the AHM possesses two phases: a phase with a spontaneously broken U(1) symmetry (the condensed or superconducting phase) and an unbroken (Coulomb) phase with a massive scalar field. In the latter phase, the theory is known as ``scalar QED''. The AHM also possesses a region near a second-order phase transition where the mass gap vanishes, so the model effectively approaches a conformal limit. Therefore, all three interesting cases -- represented by the conformal (scale-invariant), symmetric (massive fields with explicitly broken gauge symmetry), and broken (gapped phase with spontaneously broken gauge symmetry) regions -- can be studied within the same model. 

In the next section, we briefly discuss a qualitative physical picture behind the generation of the boundary current. We show that one can generally expect the existence of both conformal (quantum) and non-conformal (semiclassical) components to the boundary current. As it follows from their names, the conformal electric current is essential in a region of parameter space close to the conformal point. At the same time, the non-conformal contribution plays a dominant role in a part of the phase diagram with a mass gap in the spectrum of the matter field(s).

In Section~\ref{sec_model}, we describe the lattice model, observables, and the parameters of our lattice simulations. Then, Section~\ref {sec_numerical} is devoted to a detailed description of our numerical results. Finally, our conclusions are summarized in the last section. 

\section{Boundary current}

\subsection{Anomalous generation of the boundary current and conformal symmetry}

The scale (conformal) anomaly produces an electric current near a flat boundary in a general class of spatially bounded quantum field theories with $U(1)$ gauge symmetry~\cite{McAvity:1990we,Chu_2018ksb}. The current takes the following general form~\cite{Chu_2018ksb,Chu:2018ksb}:
\beqn
{\bs j}({\bs x}) = - \nu({\bs x}) e\, {\bs n} \times {\bs B}({\bs x})\,,
\label{eq_j_conformal}
\eeqn
where ${\bs B}$ is the strength of the magnetic field, $e = |e|$ is the elementary electric charge, and ${\bs n}$ is the outward spatial vector normal to the boundary surface. The anomalous factor
\beqn
\nu({\bs x}) \equiv \nu(x_\perp) = \frac{2 \beta(e)}{e^3} \frac{1}{x_\perp}\,,
\label{eq_nu}
\eeqn
is a function of the distance~$x_\perp$ from the boundary to the point ${\bs x} = ({x_\perp,{\bs x}_\|})$ at which the current is generated. In our notations, $x_\perp = x$ and ${\bs x}_\| = (y,z)$.

The anomalous coefficient~\eq{eq_nu} depends on the beta function $\beta$ associated with the renormalization of the electric charge. Equation~\eq{eq_j_conformal} implies that the induced electric current is tangential to the boundary and, simultaneously, normal to the axis of the magnetic field. 

The electric current at the boundary~\eq{eq_j_conformal} is proportional to the beta function $\beta$, which enters the proportionality coefficient~\eq{eq_nu}. The beta function is associated with the renormalization of electric charge:
\begin{align}
    \beta(e) = \mu \frac{\partial e(\mu)}{\partial \mu}\,,
    \label{eq_beta_function}
\end{align}
where $\mu$ denotes the energy scale at which the electric charge is measured: $e = e(\mu)$. Thus, the interactions break the conformal symmetry of the system via renormalization of the electric charge: the properties of particles interacting at different energies are no more related to each other by a simple rescaling transformation. The presence of the beta function~\eq{eq_beta_function} highlights the scale--anomalous nature of the produced electric current~\eq{eq_j_conformal}. 

The conformal anomaly is also called the scale anomaly because the conformal and scale properties of reasonably defined field theories are inherently connected to each other~\cite{Nakayama:2013is}. The same type of anomaly is also often called the trace anomaly because in the presence of the background classical electromagnetic field $F^{\mu\nu}$, the trace of energy-momentum tensor $T^{\mu\nu}$ is nonzero: $T^\mu_{\ \mu} = \beta(e)/(2e) F^{\mu\nu} F_{\mu\nu}$. The last relation is helpful for first-principle calculations of the equation of state in non-perturbative gauge theories~\cite{Boyd:1996bx}. The loss of tracelessness originates from radiative corrections, as the strength-energy tensor should otherwise be traceless in classical conformal field theories in 3+1 dimensions. More details on conformal anomalies and transport can be found in Ref.~\cite{Chernodub:2021nff}.

Our paper concentrates on scalar QED (sQED), which is much easier to simulate numerically than the usual QED with massless or light fermions. The one-loop $\beta$ function of sQED with one bosonic species coupled to a single Abelian gauge field is four times smaller compared to the conventional QED:
\beqn
\beta^{{\text{1-loop}}}_{{\text{sQED}}} = \frac{e^3}{48 \pi^2}\,,
\label{eq_beta_sQED}
\eeqn
The anomalous coefficient~\eq{eq_nu} is therefore given by the following simple expression:
\beqn
\nu^{{\text{1-loop}}}_{\mathrm{cQED}}(x_\perp) = \frac{1}{24 \pi^2} \frac{1}{ x_\perp}\,.
\eeqn

The direction of the anomalous electric current~\eq{eq_j_conformal} is tangential to the boundary and normal to the external magnetic field ${\bs B}$. Therefore, it is natural to characterize the system by calculating the total current density per unit area of the boundary. In the conformal limit, the total electric current density induced at the boundary~\eq{eq_j_conformal},
\beqn
J_{\mathrm{tot}} = \int_0^{\infty} j_\|(x_\perp)  \, d x_\perp\,,
\label{eq_int_j}
\eeqn
is a linear function of the background magnetic field:
\beqn
J_{\mathrm{tot}} = \gamma eB\,,
\eeqn
where the proportionality coefficient $\gamma$, given by the integration of Eq.~\eq{eq_j_conformal} along the normal direction $x_\perp$, diverges both in infrared and ultraviolet limits:
\beqn
\gamma^{\mathrm{th}}_{\mathrm{conf}}  = \frac{1}{24 \pi^2} \ln \frac{\lambda_{\mathrm{IR}}}{\lambda_{\mathrm{UV}}}\,.
\label{eq_gamma_conf_th}
\eeqn

In the conformal limit, the infrared cutoff~$\lambda_{\mathrm{IR}}$ in Eq.~\eq{eq_gamma_conf_th} should be of the order of the size of the system, while the ultraviolet cutoff $\lambda_{\mathrm{UV}}$ should be determined by a typical shortest scale of the material at which the continuum description of the particle's motion is no more applicable (for example, in a crystal, the value of the cutoff $\lambda_{\mathrm{UV}}$ is of the order of an interatomic distance). In our simulations, the lattice spacing naturally gives the ultraviolet scale, $\lambda_{\mathrm{UV}} = a$. For a typical lattice size $L = 32$ (given in the units of lattice spacing $a$, with more details given below), the most extended physical size along any axis is $L/2 = 16$, implying that in our simulations, the logarithmic factor in Eq.~\eq{eq_gamma_conf_th} gives a modest multiplicative correction, $\ln 16 \simeq 2.8$. 
Despite the ultraviolet divergence both in infrared and ultraviolet limits, the total anomalous current~\eq{eq_gamma_conf_th} gives a reasonable value even in the realistic systems where the ultraviolet scale, given by an interatomic distance with a typical value of a few {\AA}ngstr{\"o}ms (we take $\lambda_{\mathrm{UV}} = 1\,{\mathrm\AA} \equiv 10^{-10} \, \mathrm{m}$) is many orders of magnitude smaller than the typical crystal size (we take $\lambda_{\mathrm{IR}} = 1 \,{\mathrm{cm}}\equiv 10^{-2} \, \mathrm{m}$). Then, the logarithmic factor is $\ln (\lambda_{\mathrm{UV}}/\lambda_{\mathrm{IR}}) \simeq 18.4 \simeq 6\pi$.

\subsection{A single model for three regimes}

The anomalous boundary current~\eq{eq_j_conformal} has been found in the conformal limit where the model has no dimensionful parameters that could fix a scale. However, the conformal symmetry can be broken, and this breaking can proceed via two mechanisms: either perturbatively, via radiative corrections, or non-\-pertur\-bati\-vely, via the condensate. In our paper, we ask a natural question: can the boundary current be generated in a non-conformal phase that possesses a mass scale so that the conformal symmetry is explicitly broken? Furthermore, how does the breaking of the U(1) gauge symmetry by the condensate affects the generation of the current? In the condensate phase, both the conformal symmetry and the continuous gauge symmetry are broken simultaneously, which allows us to ask yet another question: how does the interplay between these two symmetries contribute to the generation of the anomalous current? 

In order to address all three regimes, we consider the Abelian Higgs model with the Lagrangian:
\beqn
\cL = - \frac{1}{4} F_{\mu\nu} F^{\mu\nu} + (D_{\mu} \phi)^* D^{\mu} \phi - V(\phi)\,,
\label{eq_L_Minkowski}
\eeqn
where $F_{\mu\nu} = \partial_\mu A_\nu - \partial_\nu A_\mu$ is the field strength tensor of the gauge field $A_\mu$, $D_\mu = \partial_\mu - i e A_\mu$ is the covariant derivative acting on the complex scalar field $\phi$, and 
\beqn
V(\phi) = m^2 |\phi|^2 + \lambda |\phi|^4,
\label{eq_V}
\eeqn
is the potential of the scalar field. 

The Abelian Higgs model can host two types of conformally broken regimes corresponding to the symmetric and broken phases.\footnote{In a lattice version of the Abelian Higgs model~\eq{eq_L_Minkowski} the gauge field may be a compact field, which means the presence of the third, electrically confining phase. Our paper considers a weakly coupled region ($e \ll 1$) where the confining phase is not realized.}  In the symmetric phase, the Abelian $U(1)$ symmetry is unbroken so that the physical content of the phase is the massless vector field $A_\mu$  and the massive scalar field $\phi$ with a vanishing condensate $\avr{\phi} = 0$. In the broken phase, the $U(1)$ symmetry is spontaneously broken by the nonzero condensate $\avr{\phi} \neq 0$, and both the vector field $A_\mu$ and a scalar excitation over the condensate, $\delta \phi \equiv \phi - \avr{\phi}$, acquire masses.

\begin{figure}[!thb]
\begin{center}
\begin{tabular}{cc}
\includegraphics[width=0.45\textwidth,clip=true]{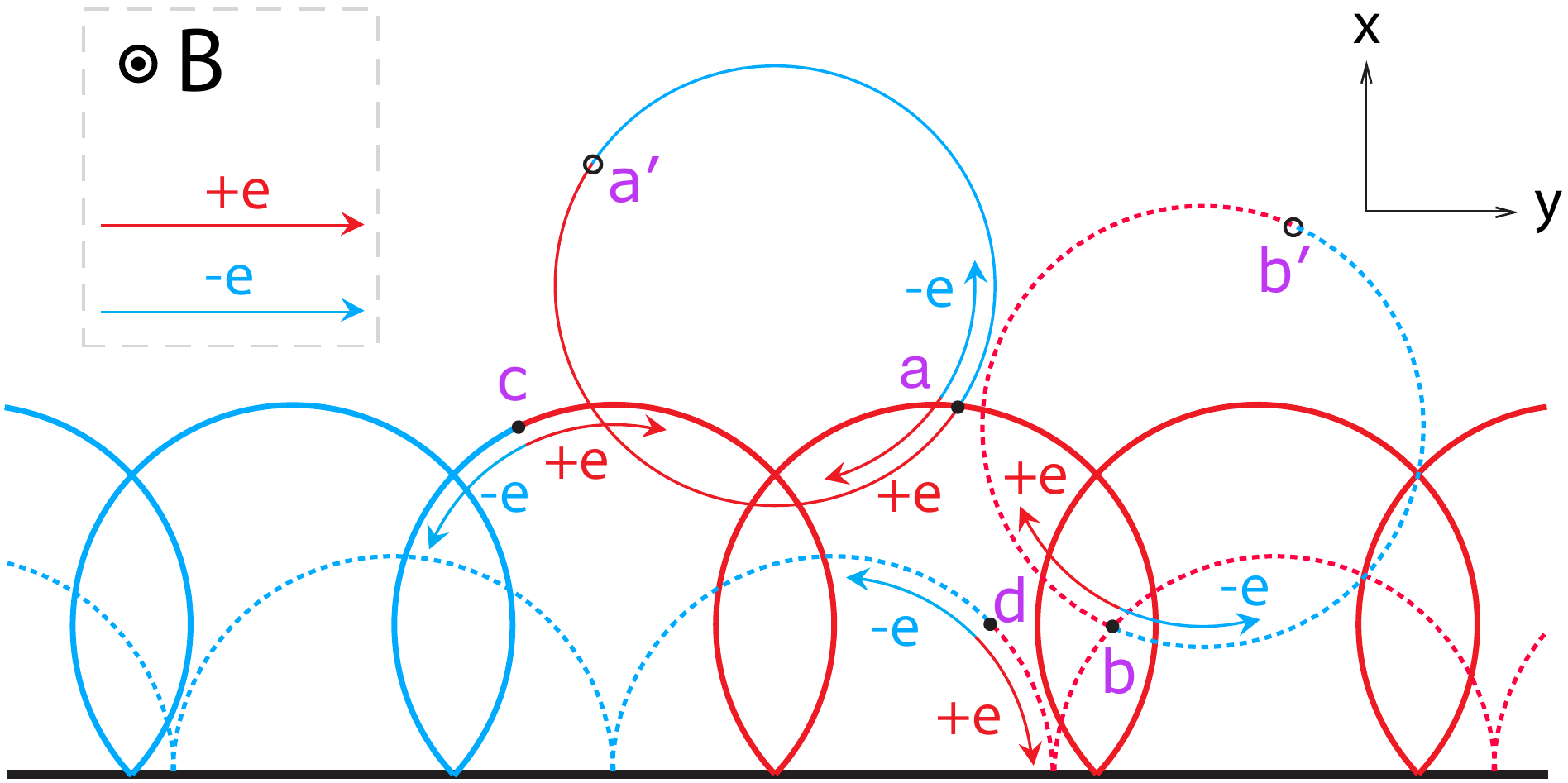} &
\hskip 5mm 
\includegraphics[width=0.45\textwidth,clip=true]{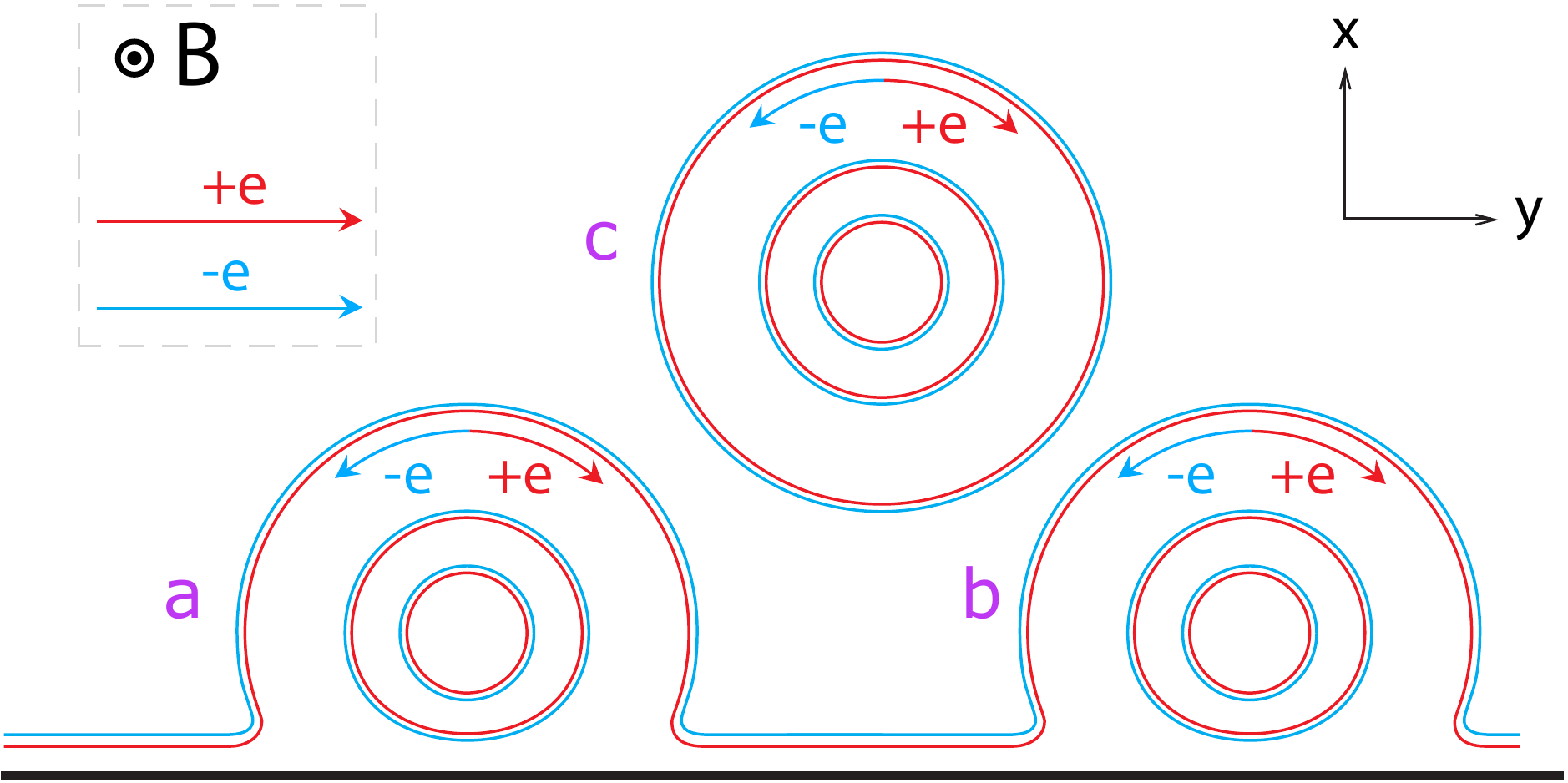} \\[2mm]
(a) & \hskip 12mm (b)
\end{tabular}
\end{center}
\vskip -4mm 
\caption{The illustration of the mechanisms of the electric current generation near the boundary in (a) the conformal regime and (b) the symmetry-broken phase. (a) Vacuum fluctuations produce particle-anti-particle pairs that follow skipping orbits without mutual annihilation, thus creating a uniform electric current along the reflective boundary. (b) Depletion of the scalar condensate acts as a well that attracts vortices to the boundary, thus enhancing the circular currents of the vortices and generating a coherent magnetization (Meissner) current along the boundary.}
\label{fig_mechanisms}
\end{figure}

\subsection{Electric current generation}
\label{sec_current_generation}

\subsubsection{Conformal current in the conformal regime}

In the conformal limit of the zero-temperature model, the quantum fluctuations produce the boundary current~\cite{McAvity:1990we,Chu_2018ksb,Chu:2018ntx}. The mechanism of the current generation is visualized in Fig.~\ref{fig_mechanisms}(a). Initially, vacuum fluctuations create particle-anti-particle pairs which, placed far from a boundary, annihilate within the time interval stipulated by the uncertainty principle. The presence of the magnetic field background does not change the fate of these virtual particles, which will, in this case, follow close circular orbits. 

However, a virtual particle-antiparticle pair created near the reflective boundary does not close its shared orbit, provided the magnetic field is directed tangentially to the boundary. As a result, the virtual particles follow skipping orbits near the wall without mutual annihilation, thus creating an electric current along the boundary (notice that particles with opposite charges move in opposite directions along the wall, thus doubling the total current). The current profile near the boundary has an infinite thickness in the normal direction to the wall because -- in the absence of any scales -- the current decays polynomially (not exponentially) at large distances from the boundary. 

The produced current has an entirely quantum origin since it originates from the vacuum at zero temperature without matter. The vanishing mass of particles facilitates the creation of their pairs from the vacuum, thus enhancing the boundary current and enforcing its long-ranged nature. The existence of the boundary current in the conformal limit has been confirmed numerically from the first principles in Ref.~\cite{Chernodub_2018ihb} to which we refer an interested reader. Below, we complete the results in the conformal point by calculations at the symmetry broken phase where the current generation is governed by an entirely different mechanism and symmetric phase, where the current generation has similar features to the conformal symmetry point.

\subsubsection{Meissner current in the broken phase}

Outside the conformal region, in the broken phase with a nonzero scalar condensate, the induced boundary current is generated by Abrikosov vortices via a simple semi-classical mechanism visualized in Fig.~\ref{fig_mechanisms}(b). An external magnetic flux parallel to the boundary creates several Abrikosov vortices equal to the number of elementary fluxes in the total magnetic flux. At the center of each vortex, the scalar condensate vanishes, and the magnetic field takes its maximum value. The radial profile of the magnetic field is shaped by a circular electric current that circumvents the vortex core in the plane normal to the core. The current is zero at the center of the vortex, takes its maximal value at the distance of the order of the penetration depth $\lambda$, and decays exponentially at longer distances. 

The scalar condensate should vanish at the boundary due to the Dirichlet condition on the scalar field. The condensate is then restored to its vacuum expectation value at a certain distance from the boundary, which is of the order of the coherence length $\xi$ (the correlation length of the scalar field). The Dirichlet boundary should then attract the Abrikosov vortices because the energy of the vortex is a monotonically rising function of the value of the scalar condensate, while at the boundary, the condensate is smaller than in bulk. Therefore the vortices tend to be ``pinned'' to the Dirichlet boundary so that the two-dimen\-sio\-nal density of the vortices is expected to be increased in the vicinity of the boundary. As a result, the circular electric currents of individual vortices cancel each other in the normal direction to the boundary while simultaneously generating a net electric current along the boundary.\footnote{One may expect that the result of our considerations is valid in type-I and type-II superconductivity regimes where vortices, respectively, attract and repel each other.} Thus, we get a boundary current that should decay exponentially along the normal to the boundary. The boundary current has a non-anomalous, semi-classical origin in the broken phase. In this semi-classical scenario, the thickness of the boundary current is set by the penetration depth of the superconductor. 

This mechanism should be contrasted to the flux barrier at the surface of realistic superconductors with a vacuum (insulator) where the vortices experience tunneling due to the energy barrier at the boundary~\cite{Tinkham2004}. Likewise, to our example above, the background magnetic field generates the Meissner current along the boundary of the superconducting samples, although the boundary conditions in these cases are different. The current is generated despite the difference in the boundary conditions (the Dirichlet boundary condition in our case vs. the Neumann condition on the charged field in the Ginzburg-Landau formulation of a superconducting condensate~\cite{Tinkham2004}).

\subsubsection{Mixed regime in the symmetric phase}

In the symmetric phase, the classical Abrikosov vortex solutions do not exist, as the penetration length is infinite (the photon is a massless particle), and no topological mechanism exists that could stabilize the classical solutions. However, the described mechanism will nevertheless work since the external magnetic field induces light (quantum) vortices which may be thought of as clumps of magnetic field shaped by the vacuum currents of a fluctuating scalar field. Such quantum vortices are often observed in the numerical simulations of scalar models~\cite{Ranft:1982hf}.

The correlation length of the scalar field sets the size of the effective scalar core of the emergent vortices. Despite their non-classical nature, the fluctuating vortex lines which would tend to concentrate at the vicinity of the boundary where the Dirichlet condition naturally suppresses the value of the scalar field, thus making it close to the vanishing value of the scalar field in the center of the vortex. The elevated density of vortex lines will induce the (again, non-anomalous) boundary current, which is normal to the magnetic field. The thickness of the boundary current is set by the correlation length of the scalar field, which is the only length scale in the theory.

\section{Lattice model}
\label{sec_model}

\subsection{Action of the model and dynamical fields}

The Euclidean lattice action of the Abelian Higgs model~\eq{eq_L_Minkowski} is given by the following formula:
\beqn
S = \beta_{\latt}\sum_x \sum_{\mu<\nu = 1}^4 \left( 1 - \cos \theta_{x,\mu\nu} \right)
+ \sum_{x} \sum_{\mu=1}^4 \left| \phi_x -  e^{i (\theta_{x\mu} + \theta^B_{x\mu})} \phi_{x+\hat\mu} \right|^2
+ \sum_{x} \left( - \kappa  \left| \phi_x \right|^2 + \lambda \left| \phi_x \right|^4 \right) \,,
\label{eq_S_AHM} 
\eeqn
where $x \equiv(x_1, x_2, x_3, x_4)$ is the Euclidean space-time coordinate, $\phi_x$ is the complex scalar field and $\theta_{x\mu}$ is the dynamical (quantum) vector gauge field. The vector gauge field $\theta^B_{\mu}$ is the background (classical) magnetic field. 

The model is characterized by the lattice spacing $a$, which corresponds to the physical length of an elementary lattice link. In a naive continuum limit, $a \to 0$, the dimensionless gauge $\theta_l$ and scalar $\phi$ fields are related to their continuum counterparts (both of the dimension of mass) via the relations $A_\mu(x) = \theta_{x,\mu}/a$ and $\phi(x) = \phi_x/a$, respectively. The physical value of the lattice spacing $a$ is usually determined by matching dimensionless lattice results to known dimensionful quantities (for example, to the mass of a particular physical excitation).

In the continuum limit ($a \to 0$), the lattice gauge coupling $\beta_\latt$ in Eq.~\eq{eq_S_AHM} is related to the bare electric charge $e$ as follows: $\beta_\latt = 1/e^2$ (here, we introduce the subscript ``latt'' in the lattice coupling ``$\beta_\latt$'' in order to discriminate it from the beta function~$\beta$). The bare couplings $\kappa$ and $\lambda$ are associated, respectively, with quadratic and quartic terms of the scalar field $\phi$ in the interaction potential~\eq{eq_V} corresponding to the last line in Lagrangian~\eq{eq_S_AHM}. 

Action~\eq{eq_S_AHM} is invariant under the Abelian gauge transformations for the gauge field, $\theta_{x\mu} \to \theta_{x\mu} + \omega_x - \omega_{x+\hat\mu}$, and the scalar field, $\phi_x \to e^{i \omega_x} \phi_x$, where $\omega_x$ is an arbitrary real-valued scalar function defined at the sites of the lattice. The background gauge field $\theta^B_{x\mu}$ is insensitive to the gauge transformations.

In the lattice model~\eq{eq_S_AHM}, the gauge field $\theta_l$ is a compact field variable because the action is invariant under the discrete shifts $\theta_l \to \theta_l + 2 \pi n_l$, where $n_l \in \Z$ is an arbitrary integer. The compactness of the gauge field automatically implies the existence of Abelian monopoles in the vacuum of the theory. The monopoles condense in the strong coupling region with $e \gtrsim 1$ (small $\beta_\latt$), and the vacuum of the model becomes confining. As we are not interested in the confining effects, we keep the lattice gauge coupling sufficiently large, thus ensuring that the monopoles rarely appear in our numerical calculations. In other words, the compactness of the lattice gauge field $\theta_l$ does not influence our results.

The numerical simulations of the present article are similar to the ones of Ref.~\cite{Chernodub_2018ihb}, albeit now we work in a different region of parameters. We use the symmetric lattice $L^4$ and the elongated lattice $L^3 \times L_x$ with $L_x \geqslant L$. Both lattices correspond to a zero-temperature theory. We use $L = 32$ and $L_x = 32, 48$ with periodic boundary conditions imposed at each direction.

In this article we simulated the model~\eq{eq_S_AHM} at the fixed parameters $\beta_{\mathrm{latt}} = 4$ and $\lambda = 10$. We generated field configurations using a Hybrid Monte Carlo algorithm~\cite{Gattringer2009,Omelyan2002} and performed simulations on Nvidia GPU cards. To achieve acceptable statistics, we used about $10^6\dots 10^7$ trajectories per each value of the background magnetic field. To accelerate calculations and reduce write operations, we accumulated mean values per every 100 trajectories only, as even in this case, our simulations generated about 1.5 TB of data. We also used binning for correct error estimations for our observables.

\subsection{Background: magnetic field and boundary wall}

We insert the reflective Dirichlet boundary for the scalar field at the middle of the lattice at $x = L_x/2$:
\beqn
\phi_x {\biggl|}_{x_2 = L_x/2} = 0\,.
\label{eq_Dirichlet_boundary}
\eeqn

The static uniform magnetic background field is introduced along the $z \equiv x_3$ axis:
\beqn
\theta^B_{x,12} = \frac{2 \pi k}{L^2}\,,
\label{eq_eB_k}
\eeqn
where other components of the field-strength tensor are taken to be zero. We use the following parameterization of the background gauge field~\cite{Bali:2011qj}:
\beqn
  \theta^B_{x,2} = \frac{2\pi k}{L_x L_y} x, \qquad
  \theta^B_{x,1}|_{x = L_x - 1} = -\frac{2\pi k}{L_y} y\,,
  \label{eq_theta_B}
\eeqn
with $\theta^B_{x,3} = \theta^B_{x,4} = 0$ and $k = 0, 1, \dots , L_x L_y/2$.

The integer number $k$ in Eq.~\eq{eq_theta_B} determines the strength~\eq{eq_eB_k} of the magnetic field. The quantization of the electromagnetic gauge field~\eq{eq_theta_B} is an essential infrared property needed to satisfy the periodic boundary conditions imposed on the gauge field:
\beqn
e^{i \theta^B_{x+L,\mu}} = e^{i \theta^B_{x,\mu}}\,.
\label{eq_periodicity}
\eeqn
On the contrary, the bound on $k$ from above is an ultraviolet feature related to the presence of the shortest physical length $a$ as the value $k$ corresponds to the number of elementary fluxes introduced on the lattice in the $xy$ plane. The maximal value of $k$ in Eq.~\eq{eq_theta_B} corresponds to the lattice half-filled with Abrikosov vortices (so that the physical vortex density is given by the ultraviolet lattice cutoff $\rho_\vort \sim a^{-2}$). We will work at weak magnetic fields $\theta^B$, far from the artificially large values of the integer number $k$. The lattice magnetic field~\eq{eq_eB_k} is related to the continuum field $B$ via the relation $\theta^B_{x,12} = eB a^2$.

All dimensionful quantities in the rest of the paper are presented in lattice units, with the lattice spacing $a$ set to unity.

\subsection{Observables: vortices and currents}

The lattice density of the Abrikosov vortices is given by the integer number~\cite{Einhorn:1977qv}
\beqn
\rho_{x,\mu\nu} = \frac{1}{2\pi} \left( l_{x,\mu} + l_{x+\hat\mu,\nu} -  l_{x+\hat\nu,\mu} - l_{x,\nu} \right)\,,
\label{eq_rho}
\eeqn
defined at every plaquette $P = P_{x,\mu\nu}$ of the lattice. The vortex density~\eq{eq_rho} is the lattice curl of the gauge-invariant link variable
\beqn
l_{x,\mu} = \arg \left[ \phi^*_x e^{i {(\theta_{x\mu} + \theta^B_{x\mu})}} \phi_{x+\hat\mu} \right]\,.
\label{eq_l}
\eeqn
The variable $\rho_P$ is equal to zero, $\rho_P=0$, if no vortex pierces plaquette $P$, and it gives $\rho_P=+1$ ($\rho_P=-1$) if a vortex (an antivortex) pierces the plaquette $P$. 

The electric current is given by the variation of the action~\eq{eq_S_AHM} with respect to the gauge field $\theta_{x\mu}$:
\beqn
j_{x\mu} = - 2 \avr{{\mathrm{Im}} \left[ \phi^*_x e^{i {(\theta_{x\mu} + \theta^B_{x\mu})}} \phi_{x+\hat\mu} \right]}\,.
\label{eq_j}
\eeqn

\section{Numerical results}
\label{sec_numerical}

This section reports the results of the first-principle numerical simulations on the boundary current in both the symmetric and broken phases of the Abelian Higgs model defined in Section~\ref{sec_model} and compares them with our theoretical expectations described earlier in Section~\ref{sec_current_generation}.

\subsection{Phase transition}

First, we examine the symmetric phase far from the conformal regime. To eliminate the presence of Abelian monopoles and make the gauge field behave similarly to the continuum theory~\eq{eq_L_Minkowski}, we set the large gauge coupling to a fixed value of $\beta = 4$. In addition, we choose a small value for the quartic constant, namely $\lambda = 0.01$, which guarantees the absence of a second-order phase transition at any value of the two-point coupling~ $\kappa$. We work at the lattices $32^4$ and $32^3 \times 48$. In the middle of the lattice, at $y = 16$, we put the Dirichlet boundary~\eq{eq_Dirichlet_boundary} to simulate a reflective wall.

\begin{figure}[!thb]
\begin{center}
\includegraphics[width=0.5\textwidth,clip=true]{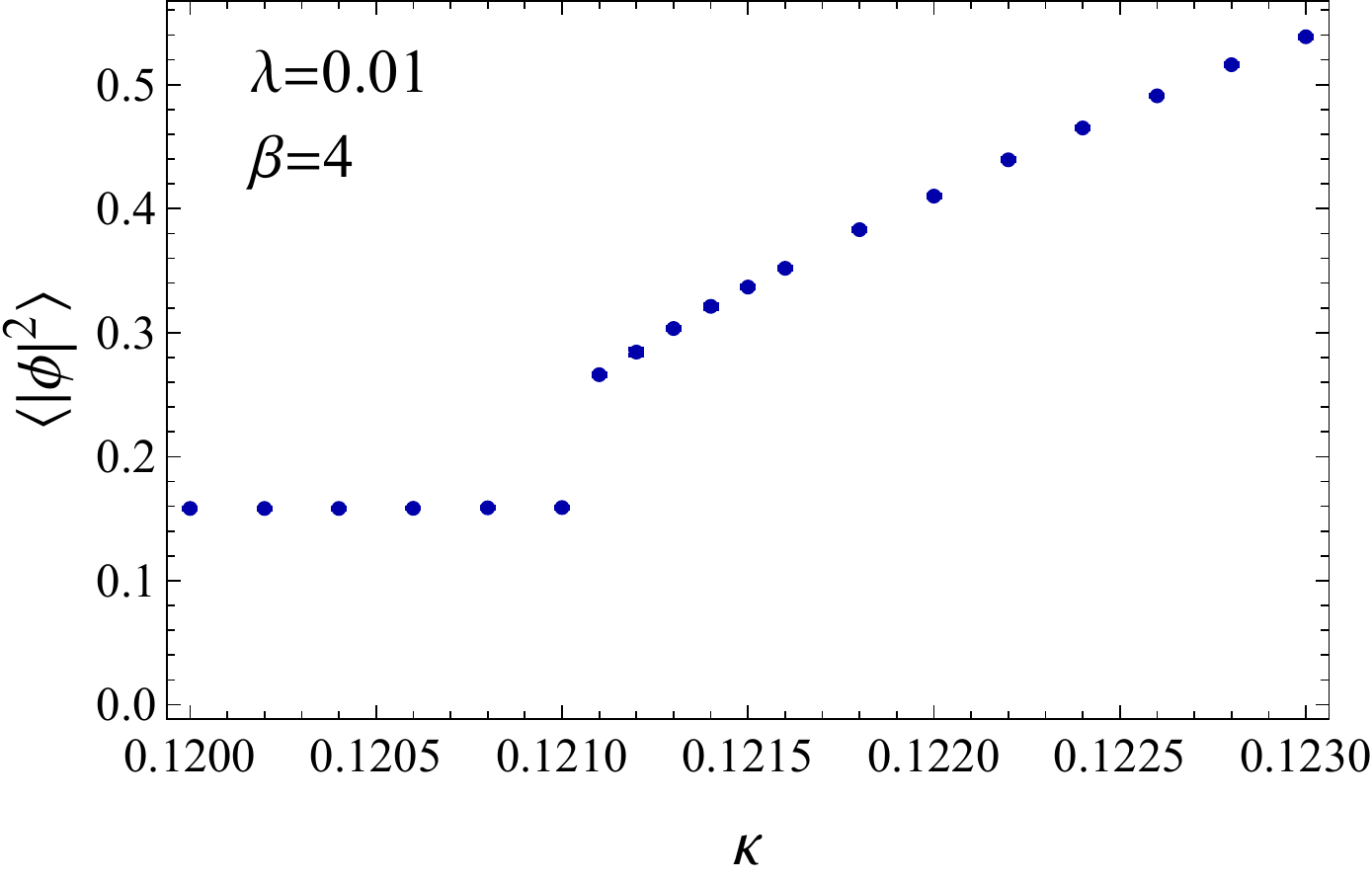}
\end{center}
\vskip -4mm 
\caption{The (unrenormalized) quadratic fluctuations of the scalar field as the function of the lattice hopping parameter~$\kappa$ in the vicinity of the first-order phase transition at fixed quartic coupling $\lambda = 0.01$ and the gauge coupling $\beta = 4$ at $32^4$ lattice at a vanishing magnetic field, $B=0$. The discontinuity in the quadratic fluctuations marks the first-order phase transition.}
\label{fig_phi2_beta4}
\end{figure}
The behavior of the quadratic fluctuations of the scalar field $\avr{|\phi|^2}$ as the function of the hopping coupling~$\kappa$, Fig.~\ref{fig_phi2_beta4}, demonstrates the existence of a first-order phase transition at $\kappa_c \simeq 0.121$ with the symmetric (broken) phase at smaller $\kappa < \kappa_c$ (larger $\kappa > \kappa_v$) values of the coupling, separated by a discontinuity in $\avr{|\phi|^2}$. 

Therefore in the symmetric and broken phases, we choose the following set of lattice parameters, respectively:
\begin{align}
    \beta & = 4, \quad \lambda = 0.01, \quad \kappa = 0.12\quad \mbox{[symmetric]}\,,
\label{eq_par_symmetric} \\
    \beta & = 4, \quad \lambda = 0.01, \quad \kappa = 0.2\phantom{0}\quad \mbox{[broken]}\,.
\label{eq_par_broken}
\end{align}

\begin{figure}[!thb]
\begin{center}
\includegraphics[width=0.5\textwidth,clip=true]{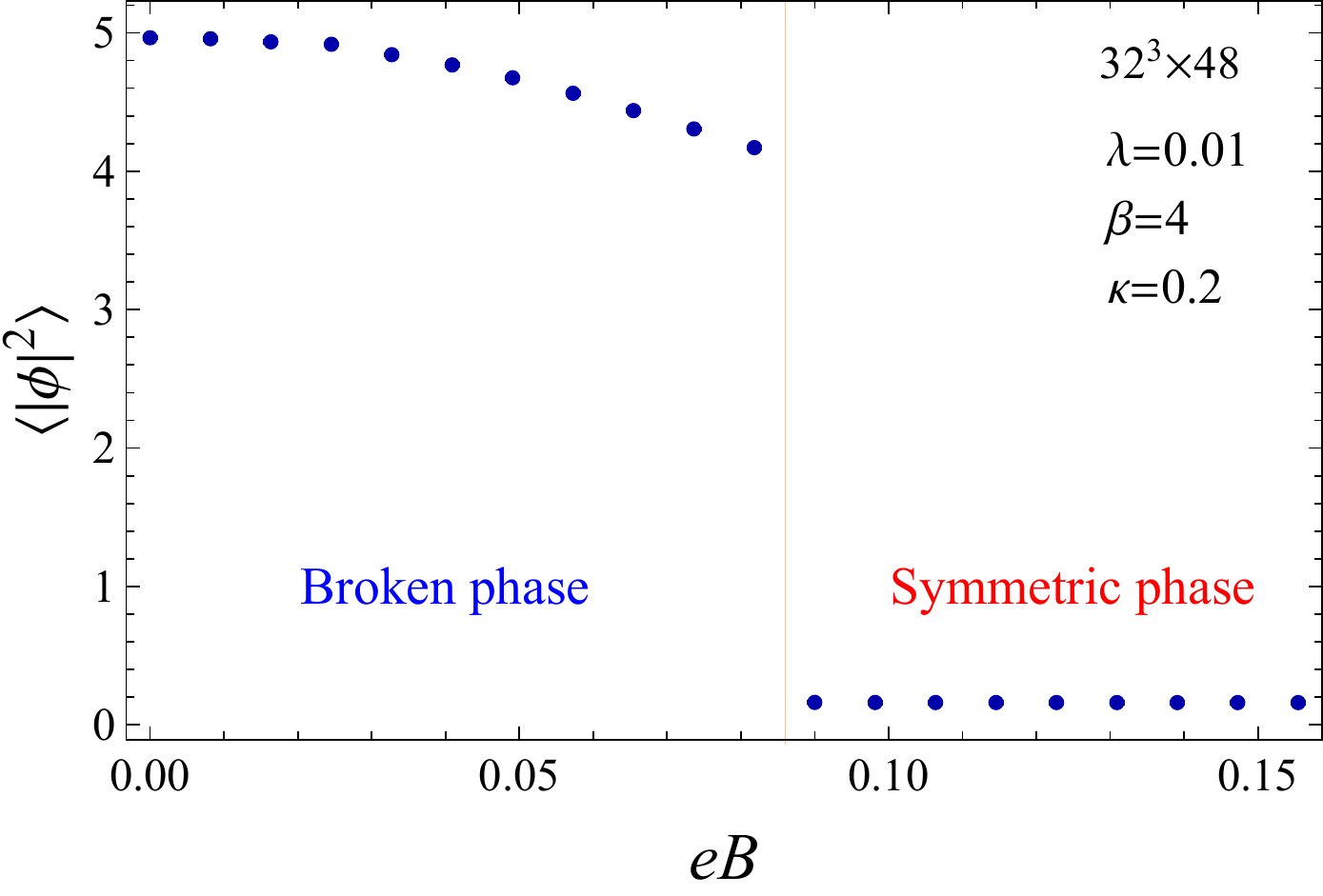}
\end{center}
\vskip -4mm 
\caption{The quadratic fluctuations of the scalar field as the function of the magnetic field strength $eB$ at fixed couplings~\eq{eq_par_broken} corresponding to the broken phase at $32^3\times 48$ lattice. The thin red vertical line approximately marks the position of the critical magnetic field, which corresponds to the first-order phase transition from the broken phase to the symmetric phase.}
\label{fig_phi2_vs_eB_48}
\end{figure}

The presence of the background magnetic field can also influence the position of the phase transition point by restoring the symmetric (normal) phase from the broken (superconducting) phase. At the chosen point of parameters~\eq{eq_par_broken}, this transition happens at a relatively large value of the magnetic field, 
\begin{align}
    eB_c \simeq 0.87\,, 
\label{eq_Bc}
\end{align}
which is seen as a discontinuity in the expectation value $\avr{|\phi|^2}$, Fig.~\ref{fig_phi2_vs_eB_48}. Therefore, we consider the effect of the current generation with the fields lower than the critical value, $B < B_c$. On the contrary, the model expectedly remains in the symmetric phase at all considered values of the background magnetic field at our set of parameters~\eq{eq_par_symmetric}. Below, we study these phases in more detail.

\subsection{Symmetric phase}

\begin{figure}[!thb]
\begin{center}
\includegraphics[width=0.55\textwidth,clip=true]{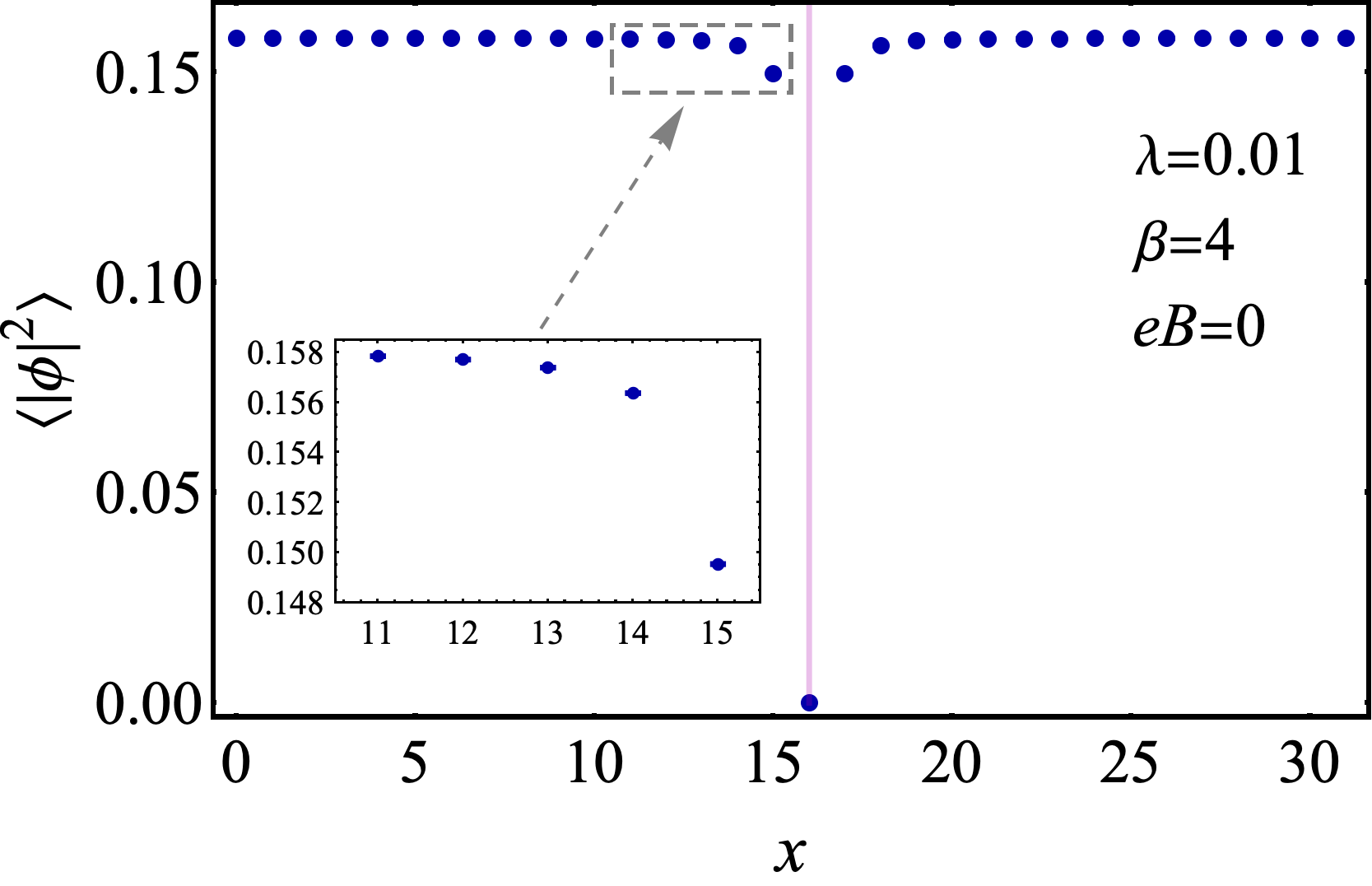}
\end{center}
\vskip -4mm 
\caption{Symmetric phase: The mean value of the scalar field. The light magenta line marks the position of the boundary. Simulations are performed at the parameter set~\eq{eq_par_symmetric}.}
\label{fig_phi2_mean}
\end{figure}

First, we consider the symmetric phase represented by the set of parameters~\eq{eq_par_symmetric}. In Fig.~\ref{fig_phi2_mean}, we show the expectation value of the scalar field fluctuations $\avr{|\phi|^2}$ at a zero magnetic field. As expected, the fluctuations vanish at the boundary and recover quickly to the bulk value in a few lattice spacings from the boundary wall. This picture is practically independent of the strength of the background magnetic field. 

\begin{figure}[!thb]
\begin{center}
\includegraphics[width=0.5\textwidth,clip=true]{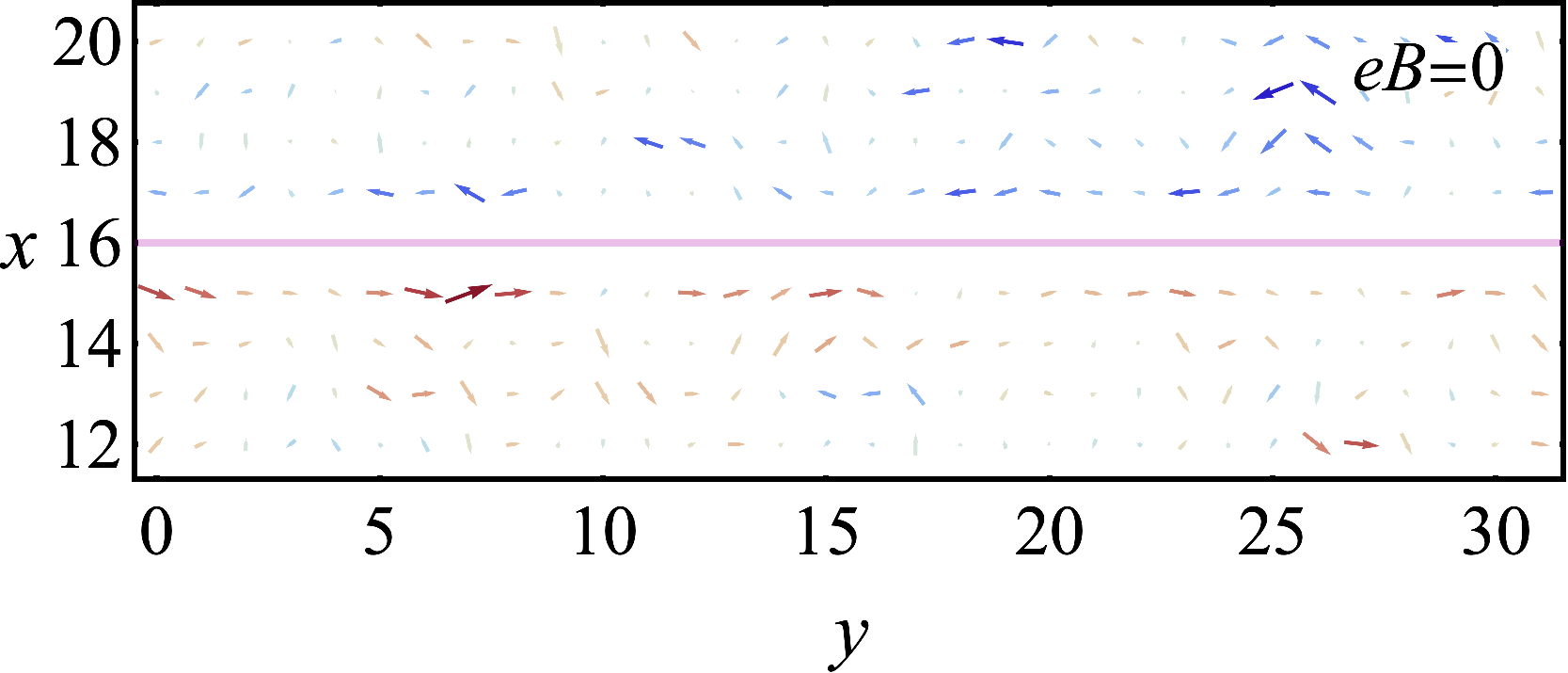}\\[-5mm]
\includegraphics[width=0.5\textwidth,clip=true]{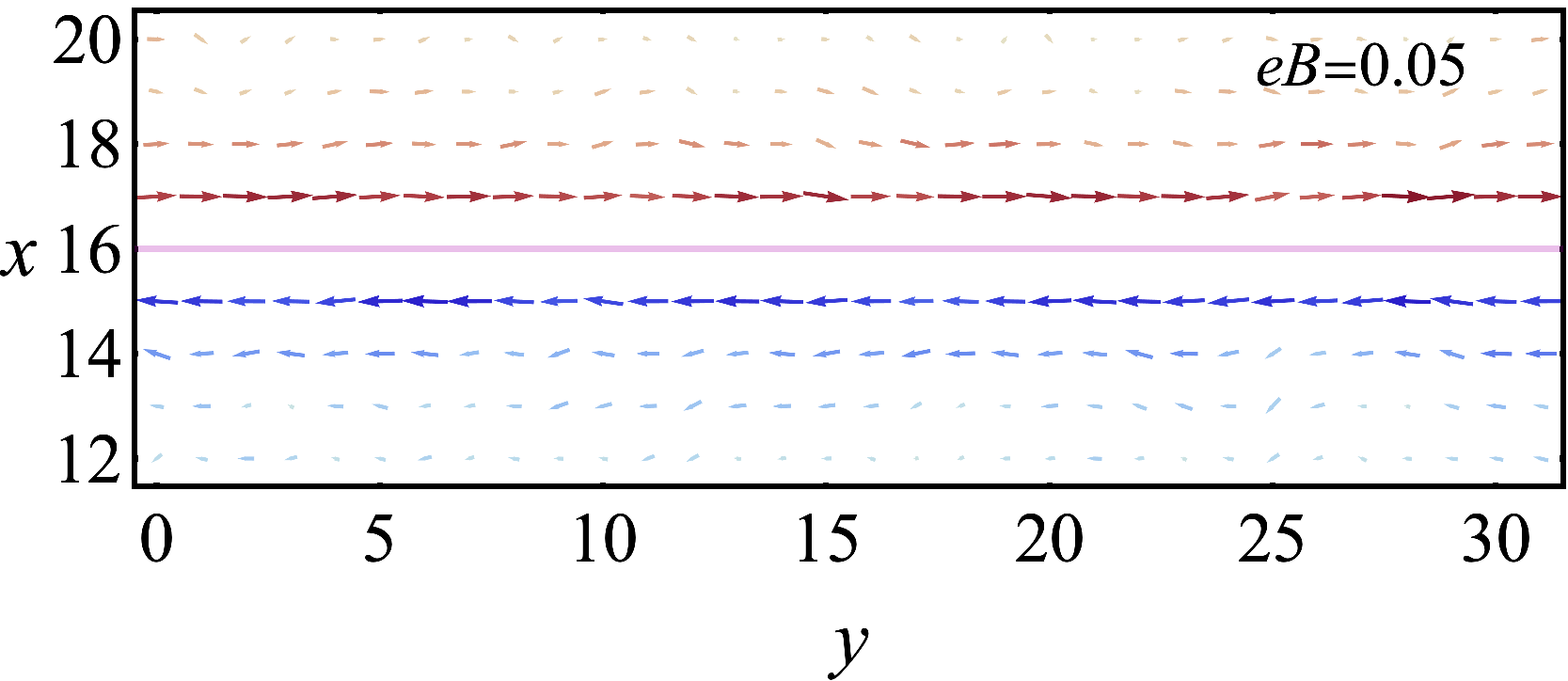}\\[-5mm]
\includegraphics[width=0.5\textwidth,clip=true]{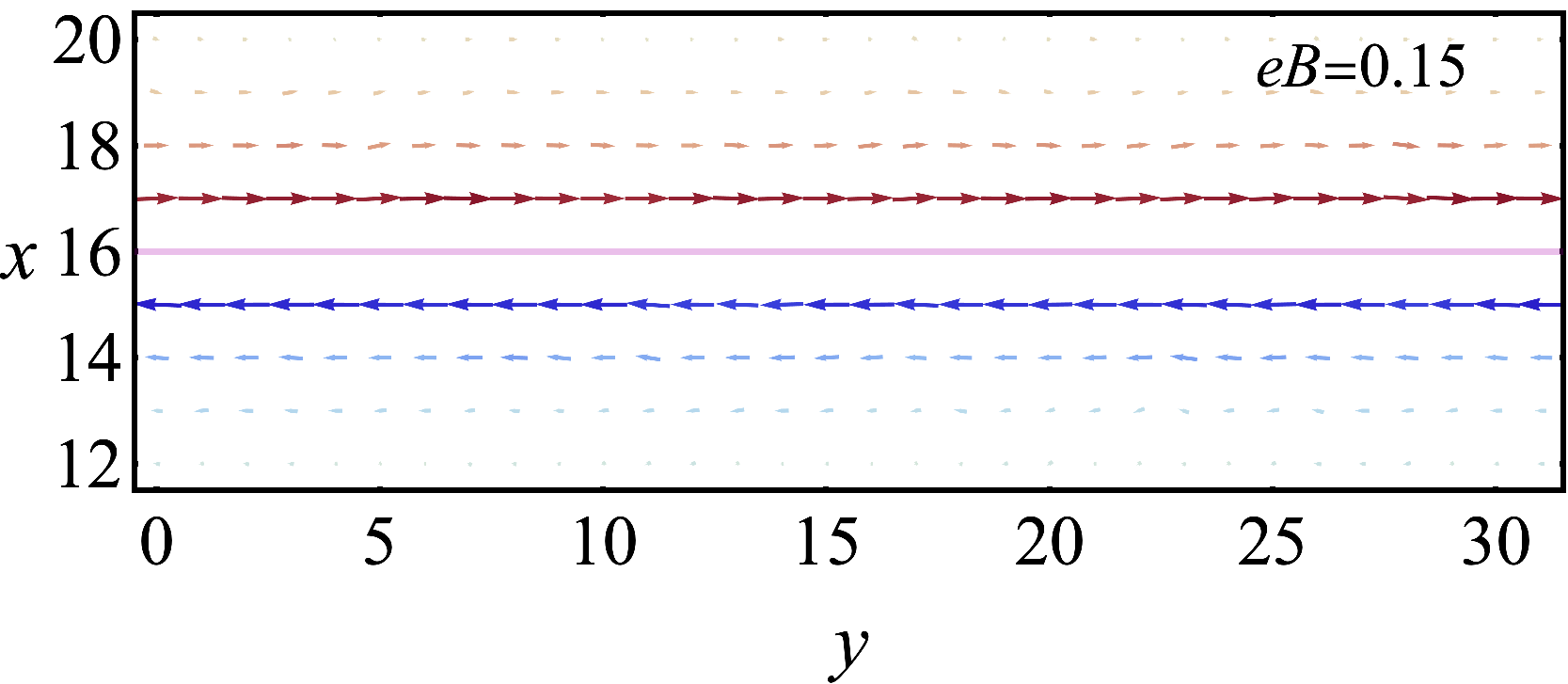}
\end{center}
\vskip -4mm 
\caption{Symmetric phase~\eq{eq_par_symmetric}: The (normalized) electric current in the $xy$ plane in the vicinity of the wall (positioned at $x=16$ and directed along the $y$ axis) at three values of increasing magnetic field. The current is calculated with a moderate number of lattice field configurations which allow us to highlight the magnitude of its fluctuations. The color of the arrows corresponds to the sign of the $y$-component of the current.}
\label{fig_j_2d}
\end{figure}

Figure~\ref{fig_j_2d} shows the expectation value of the electric current~\eq{eq_j}, revealing the presence of the electric current perpendicular to the external magnetic field. We calculated by averaging the local current over a moderate number of configurations to highlight the magnitude of quantum fluctuations. When the magnetic field is zero, the current appears as a small random vector (top panel in Fig.~\ref{fig_j_2d}), with a vanishing expectation value if averaged over a sufficiently large number of configurations. However, as a nonzero magnetic field is applied tangentially to the boundary, the presence of the boundary electric currents becomes evident. The currents flow opposite directions at opposite sides of the boundary (two lower panels in Fig.~\ref{fig_j_2d}). Increasing the strength of the magnetic field reduces the relative randomness of the generated boundary current while amplifying its intensity.

\begin{figure}[!thb]
\begin{center}
\includegraphics[width=0.5\textwidth,clip=true]{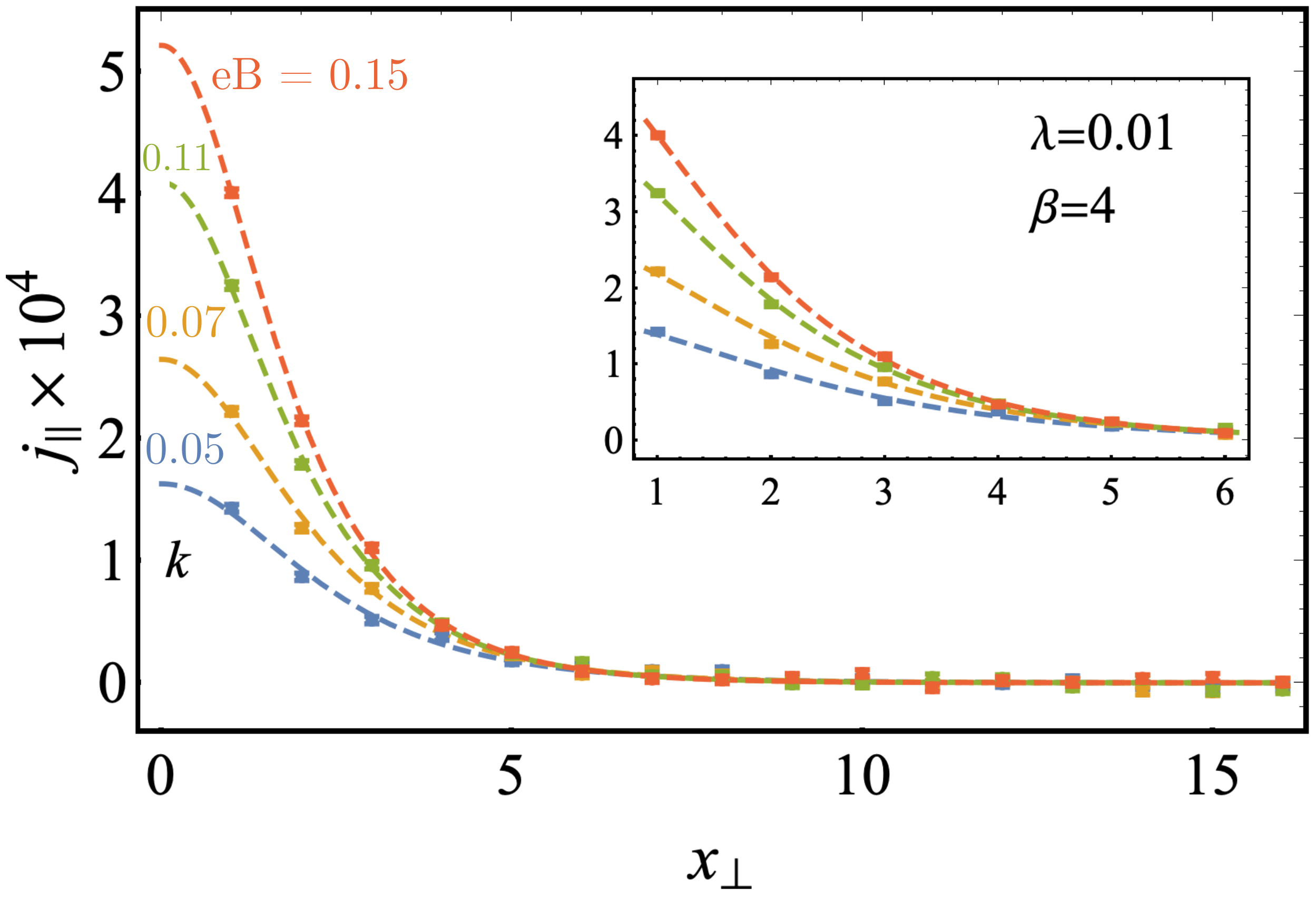}
\end{center}
\vskip -4mm 
\caption{Symmetric phase~\eq{eq_par_symmetric}: The local electric current $j_y$ at the function of the distance $x_\perp$ from the boundary at various values of the magnetic field $eB$, Eq.~\eq{eq_eB_k}. The dashed lines correspond to the best fits by function~\eq{eq_j_fit}. The inset shows a zoom-in on the region close to the wall.}
\label{fig_j_x}
\end{figure}

In Fig.~\ref{fig_j_x}, we show the local density of the boundary current $j_y$ as the function of the distance from the wall~$x_\perp$. We consider only positive values of $x_\perp$ since the current is perfectly anti-symmetric with respect to the inversion $x_\perp \to - x_\perp$. It turns out that the dependence of the boundary current on the distance from the wall $x_\perp$ can be excellently (with $\chi^2/d.o.f. \lesssim 1$) described by the following fitting function:
\beqn
j^{\mathrm{fit}}_y(x_\perp) =  \frac{2 M J_{\mathrm{tot}}}{\pi \cosh(M x_\perp)} \mathrm{sign}\, x_\perp\,,
\label{eq_j_fit}
\eeqn
with two fitting parameters: the mass parameter $M$ and the total current $J_{\mathrm{tot}}$. 

The mass parameter $M$ controls the thickness of the current in the normal direction, $\lambda_\perp = M^{-1}$, an essential feature of the current profile in the broken phase. For example, an attempt to describe the data with a conformal-like fit $j^{\mathrm{fit}}_y \sim 1/|x_\perp|$ with either infrared or ultraviolet cutoffs (or even with both) gives us unreliable results with poor qualities of the fits. The last fact is a natural consequence of the absence of conformal symmetry in a deep symmetric phase~\eq{eq_par_symmetric} which calls us to introduce the mass parameter $M$, which breaks the conformal symmetry. The fitting parameter $J_{\mathrm{tot}}$ corresponds to the total current, obtained by integration of the current~\eq{eq_j_fit} along the whole transverse direction~\eq{eq_int_j}. The total current is a finite quantity, both in infrared and ultraviolet limits. 

\begin{figure}[!thb]
\begin{center}
\begin{tabular}{cc}
\includegraphics[width=0.45\textwidth,clip=true]{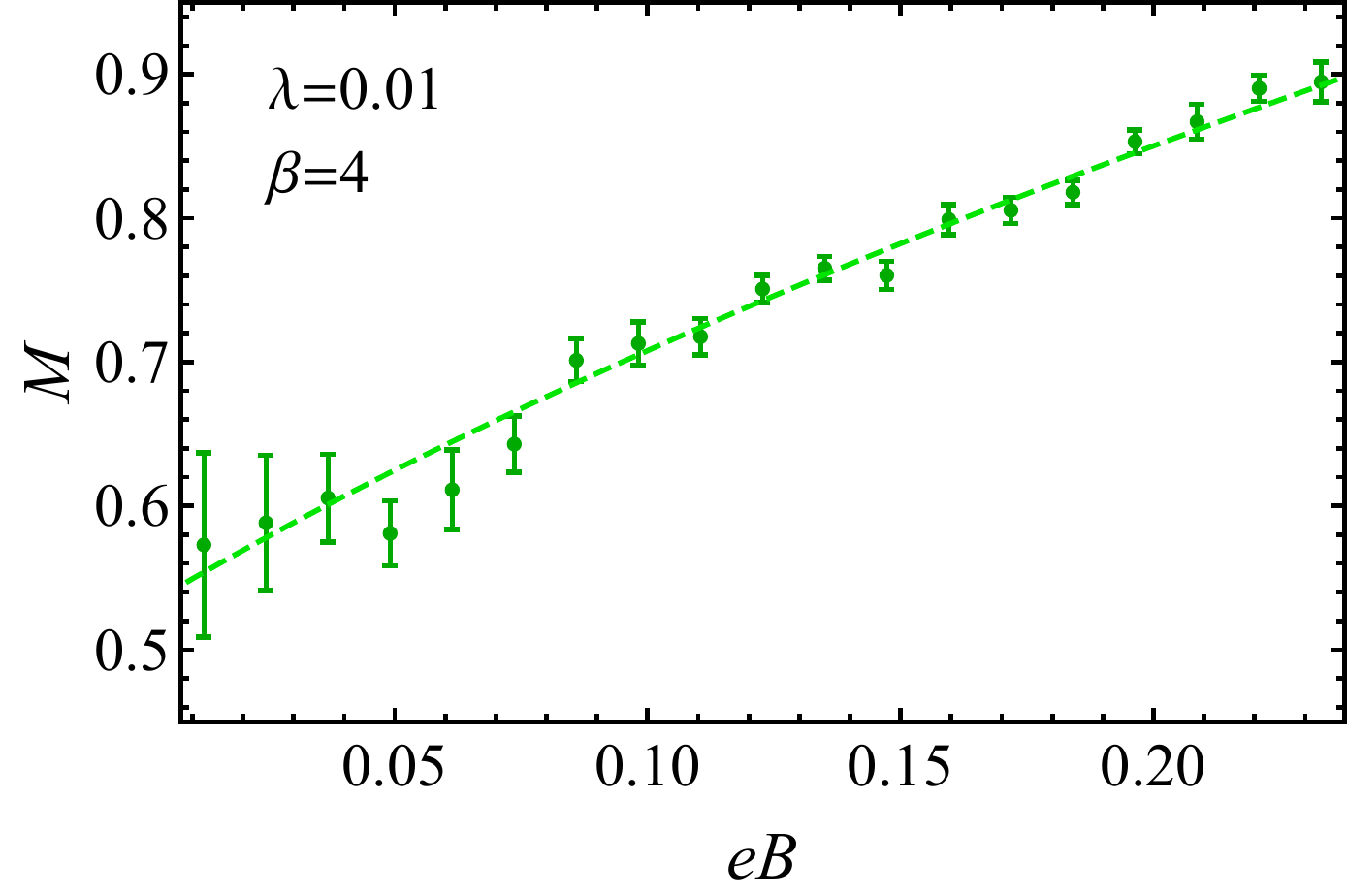} &
\hskip 5mm 
\includegraphics[width=0.45\textwidth,clip=true]{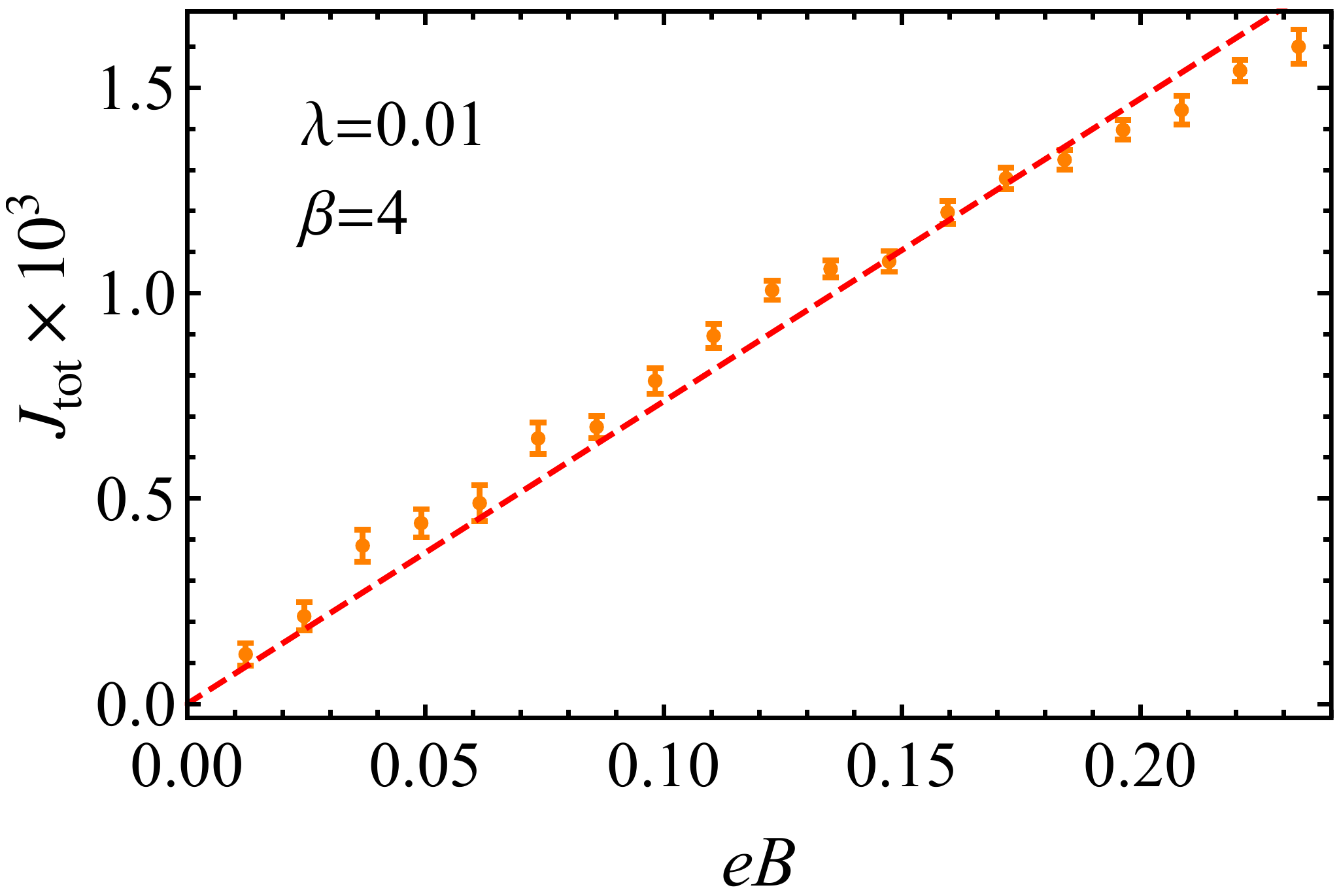} \\[1mm]
\hskip 11mm (a) & \hskip 17mm (b)
\end{tabular}
\end{center}
\vskip -4mm 
\caption{Symmetric phase~\eq{eq_par_symmetric}: (a) The screening mass factor~$M$ of the current density near the boundary~\eq{eq_j_fit} vs. the magnetic field strength. The dashed line represents the best fit~\eq{eq_M_fit}. (b) The total electric current generated at the boundary vs. the external magnetic field. The line represents the best fit by the linear function~\eq{eq_j_tot_symm}.}
\label{fig_M_J}
\end{figure}

The localization of the electric current in the vicinity of the boundary is determined by the value of the mass of the fit~\eq{eq_j_fit}, which is shown in Fig.~\ref{fig_M_J}(a) as the function of the magnetic field strength~$eB$. The localization mass $M$ increases with the background magnetic field, which is not an unusual effect since masses of elementary electrically charged scalar excitations should be rising, in a relativistic free field theory, as $M (eB) = \sqrt{m^2 + |eB|}$. The last property suggests us the form of the fit of the numerical data:
\beqn
M^{\mathrm{fit}}(eB) = \sqrt{M_0^2 + g |eB|}\,,
\label{eq_M_fit}
\eeqn
where the factor $g$ takes into account the fact that the particles are interacting, and the increase of magnetic field drives the system deeper into the symmetric phase. The best fit, shown in Fig.~\ref{fig_M_J}(a) by the dashed line, gives us the parameters $M_0 = 0.53(1)$ and $g = 2.2(1)$. Thus, the stronger the magnetic field, the thinner the current density distribution at the wall. 

In Fig.~\ref{fig_M_J}(b), we show the total boundary current as the function of the magnetic field. In a qualitative agreement with the general linear-response arguments~\eq{eq_j_conformal}, we find that the total current, similarly to the local current density, is a linear function of the background magnetic field:
\beqn
J^{\mathrm{fit}}_{\mathrm{tot}}= \gamma \, eB\,.
\label{eq_j_tot_symm}
\eeqn
Function~\eq{eq_j_tot_symm} matches our numerical data very well ($\chi^2/d.o.f. \sim 1.$), as it seen in Fig.~\ref{fig_M_J}(b).

The best fit gives us for the proportionality coefficient $\gamma = 0.0074(1)$ which matches well the order of the theoretical coefficient~\eq{eq_gamma_conf_th} as, theoretically, we expect to obtain $\gamma^{\mathrm{th}} = 1/(24 \pi^2) \approx 0.0042$. Formally, if it is the conformal anomaly that was responsible for the generation of the boundary current, then the relation between the infrared and ultraviolet cutoffs in Eq.~\eq{eq_gamma_conf_th} would be quite reasonable as well: $\lambda_{\mathrm{IR}} \approx 6 \lambda_{\mathrm{UV}}$.

\begin{figure}[!thb]
\begin{center}
\includegraphics[width=0.5\textwidth,clip=true]{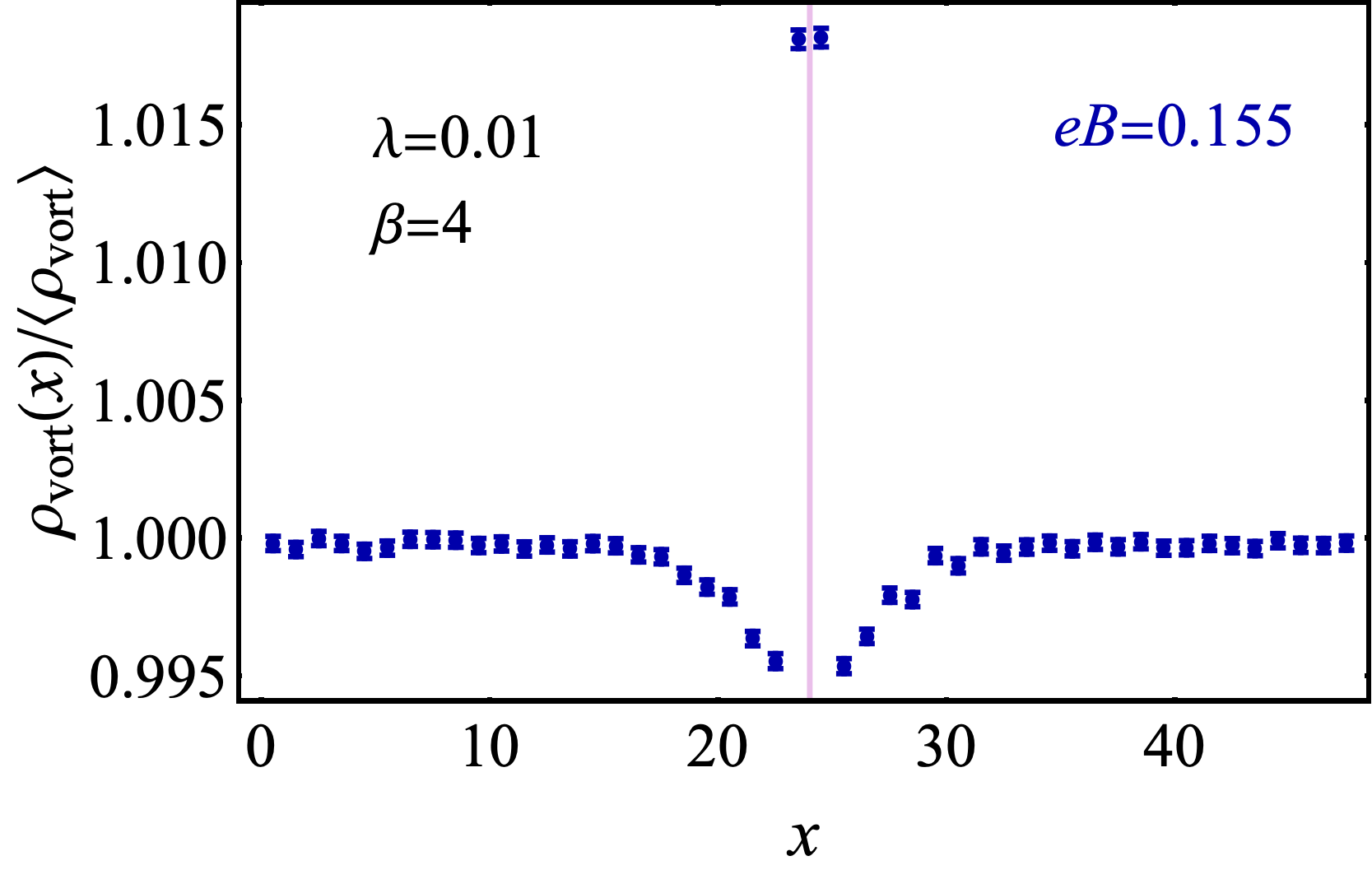}
\end{center}
\vskip -4mm 
\caption{Symmetric phase~\eq{eq_par_symmetric}: The (normalized) local vortex density as the function of the distance to the wall at the $32^3 \times 48$ lattice with an extended spatial direction $L_x = 48$.}
\label{fig_vortex_mean}
\end{figure}

The conformal anomaly cannot contribute much to the boundary current in a deep symmetric phase due to a large mass gap. In order to check the alternative, vortex-based semi-classical mechanism discussed in the introduction, in Fig.~\ref{fig_vortex_mean}, we plot the local vortex density~\eq{eq_rho} as the function of the distance to the boundary. We normalize the local vortex density to the mean vortex density on the whole lattice. The numerical data is consistent with the vortex mechanism of the current generation represented in Fig.~\ref{fig_mechanisms}(b): the density of the vortices is elevated close to the boundary, implying that the circular electric currents that circumvent the vortex cores are generating the boundary current. Figure~\ref{fig_vortex_mean} also demonstrates the existence of a repulsion force between the vortices as the boundary layer of the vortices repels the vortices from the bulk, causing a noticeable drop in the vortex density deeper in bulk.

In other words, vortices are attracted to the wall causing a small, of the order of one percent, excess of the vortex density at the wall. In the symmetric phase, the quantum vortices repel each other similarly to the type-II superconductor. Therefore, some of the vortices are concentrated precisely at the position of the wall, making the vortex density elevated exactly at the boundary. Those boundary-pinned vertices are repelling the other vortices that emerge from the bulk, thus causing the drop in the vortex density near, but not strictly at, the position of the wall. This mechanism, highlighted qualitatively in Fig.~\ref{fig_mechanisms}(b), is responsible for the physical picture observed in Fig.~\ref{fig_vortex_mean}.

We finish the study of the current in the symmetric phase by addressing the question whether the near-boundary behaviour of the current, determined numerically by the fitting function~\eq{eq_j_fit}, related to the conformal behaviour, predicted analytically by Eqs.~\eq{eq_j_conformal} and \eq{eq_nu}. The naive zero mass limit, $M \to 0$, does not recover the conformal $1/x_\perp$ behavior from the functional form $1/\cosh(M x_\perp)$. Our attempts to fit the data by the function~\eq{eq_j_fit} with $\sinh(M x_\perp)$ instead of $\cosh(M x_\perp)$ lead to inconsistent results with a poor quality of fit. In detail, we found $\mathrm{d.o.f} \simeq 10\dots 20$ for the sinh-like behaviour in comparison with good $\mathrm{d.o.f} \simeq 0.5\dots 1.5$ for the cosh-fit function~\eq{eq_j_fit}. These results imply that the point of the coupling space considered in this article~\eq{eq_par_symmetric} lies deeply in the symmetric phase which does not allow for this simple continuation to the conformal point~\eq{eq_j_conformal}.

\subsection{Broken phase}

Now we repeat all our calculations in the broken phase given by the initial set of parameters~\eq{eq_par_broken} at a larger lattice volume $32^3 \times 48$, where the longer lattice size extends one of the directions of the boundary, $L_x = 48$. This theory corresponds to zero temperature.

\begin{figure}[!thb]
\begin{center}
\begin{tabular}{cc}
\includegraphics[width=0.45\textwidth,clip=true]{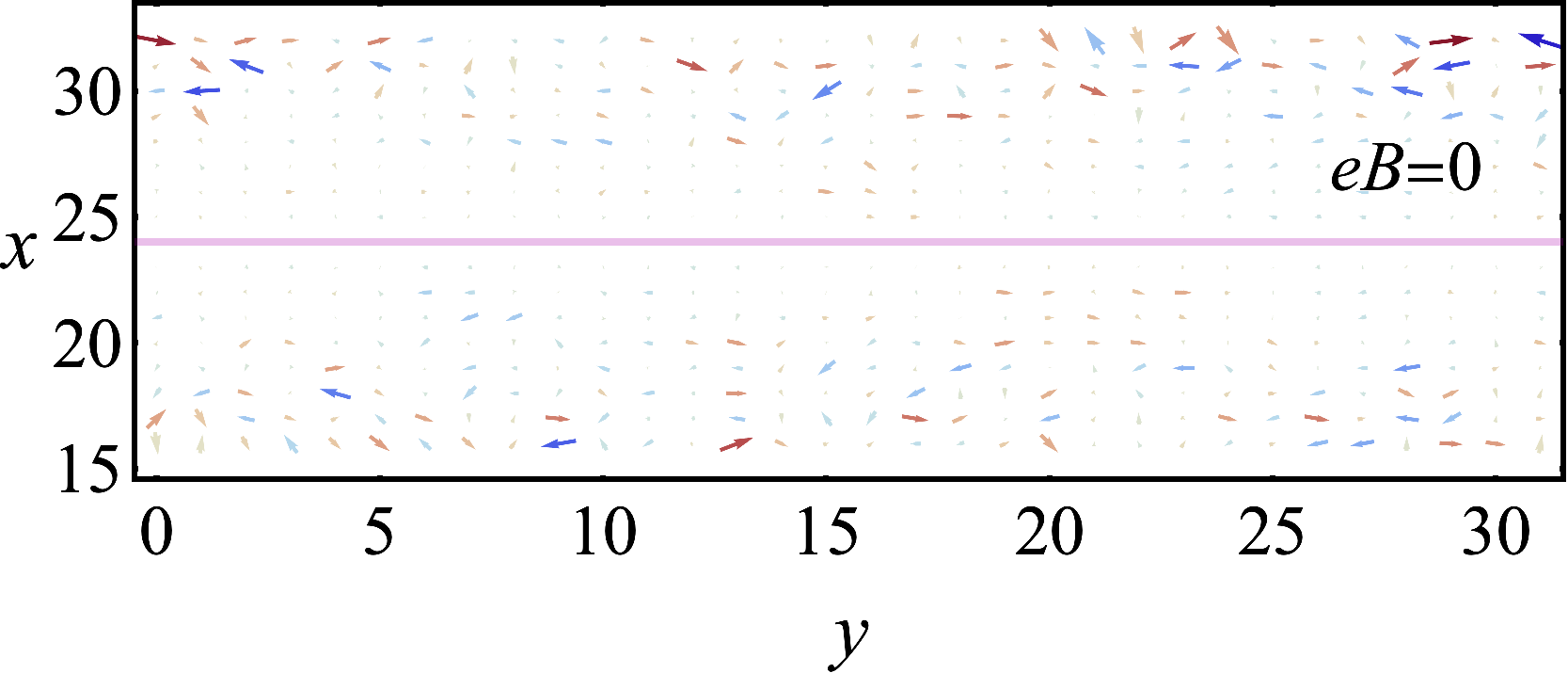} &
\hskip 5mm 
\includegraphics[width=0.45\textwidth,clip=true]{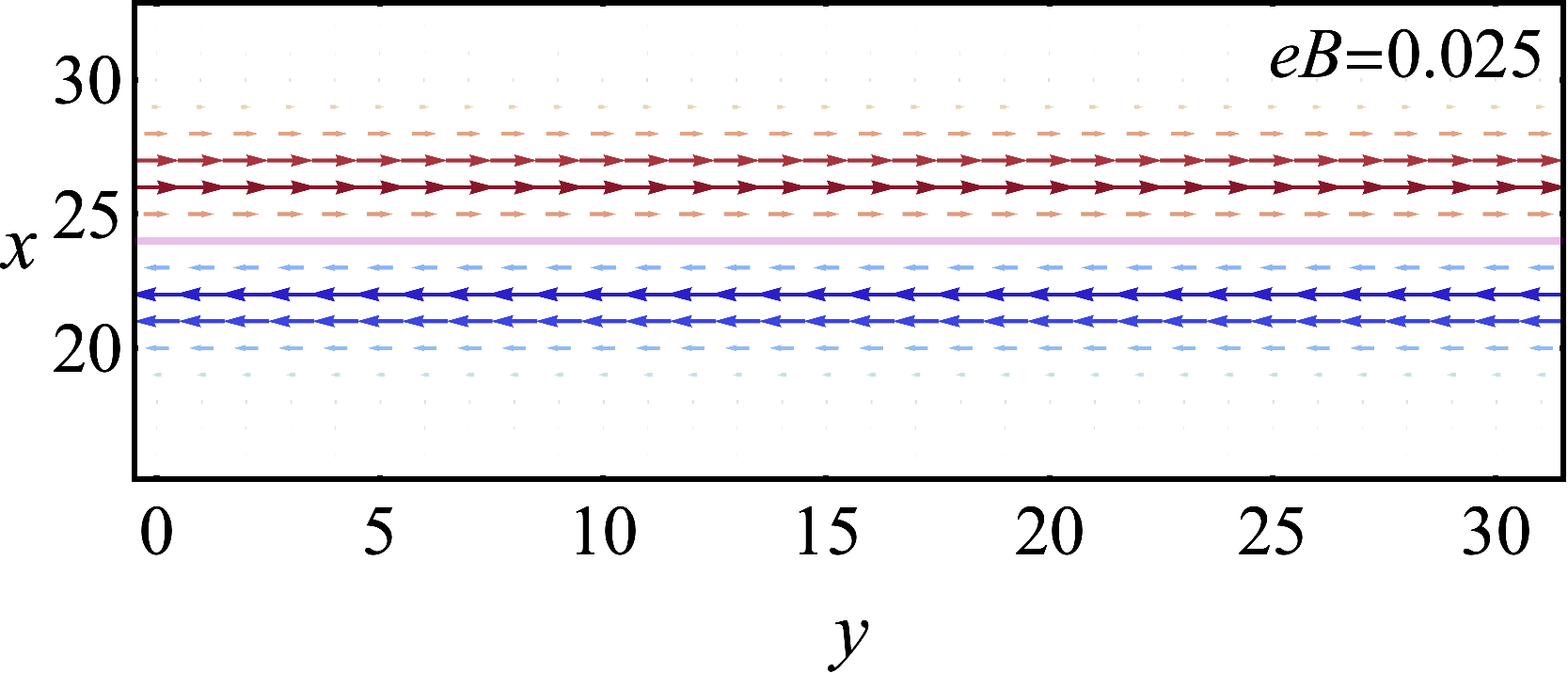} \\[2mm]
\includegraphics[width=0.45\textwidth,clip=true]{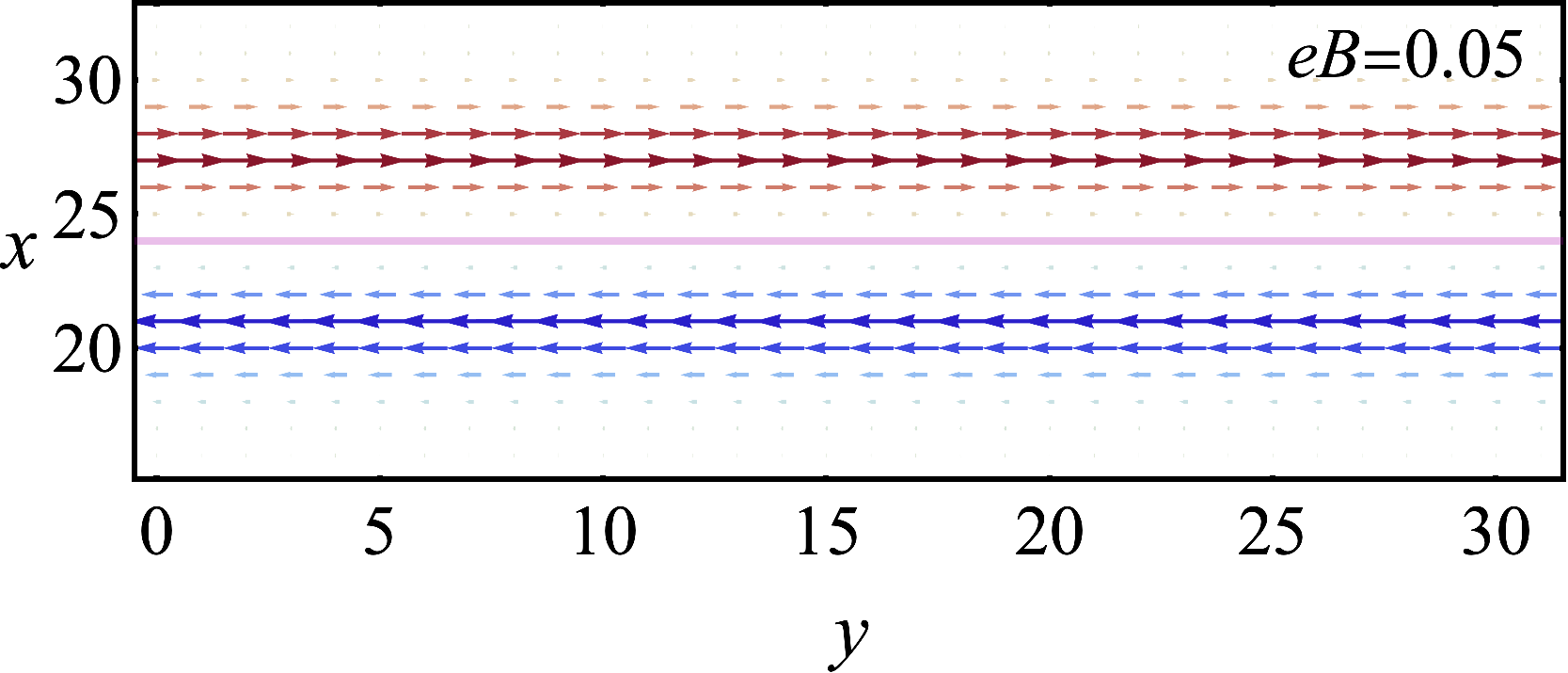} &
\hskip 5mm 
\includegraphics[width=0.45\textwidth,clip=true]{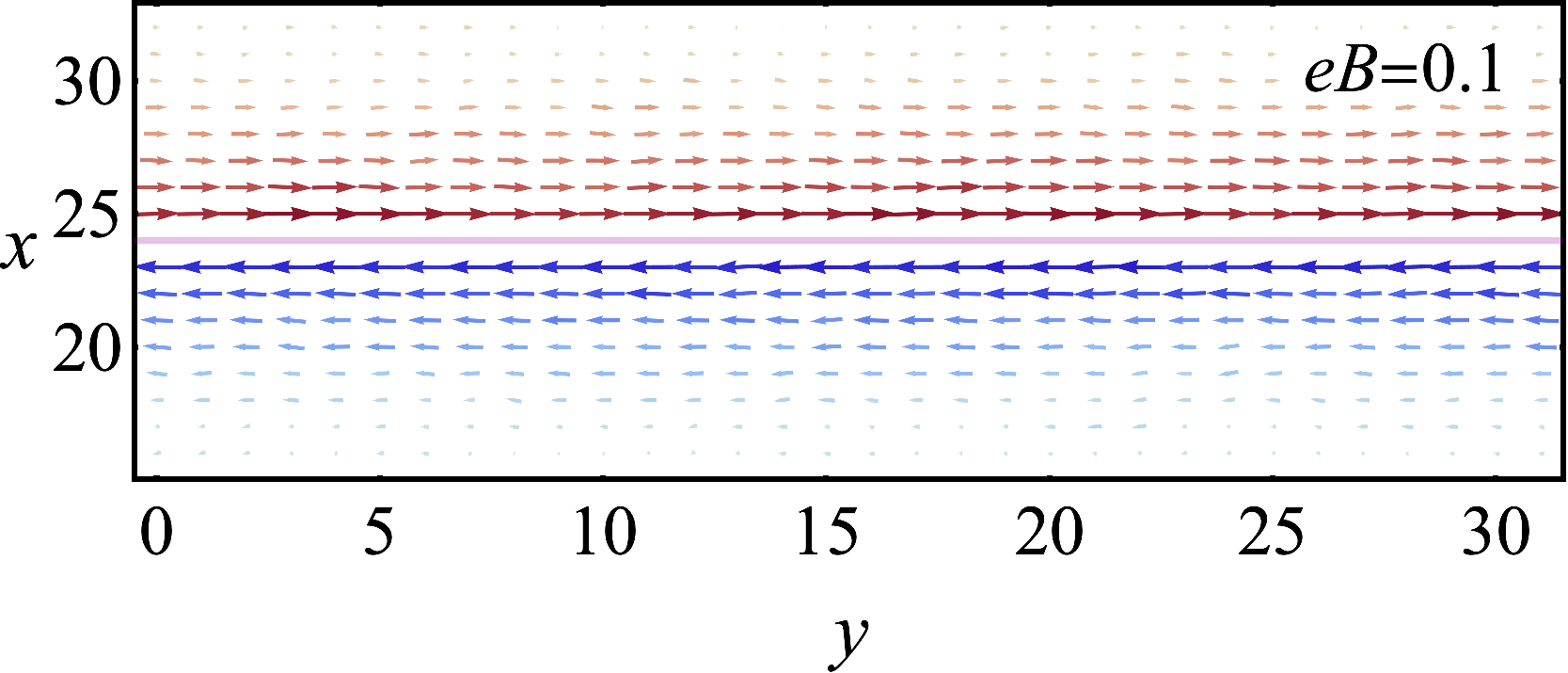}
\end{tabular}
\end{center}
\vskip -4mm 
\caption{Broken phase~\eq{eq_par_broken}: The (normalized) electric currents generated close to the wall as in Fig.~\ref{fig_j_2d} at various values of the background magnetic field $B$. Notice that the lowest panel corresponds to the symmetry restored phase at $B > B_c$ with the critical magnetic field $B_c$ given in Eq.~\eq{eq_Bc}.}
\label{fig_j_2d_48}
\end{figure}

First, we visualize in Fig.~\ref{fig_j_2d_48} the electric current generated near the wall at increasing values of the background magnetic field. While the produced electric current has the same direction as in the symmetric phase, Fig.~\ref{fig_j_2d}, one immediately notices the difference between the broken phase and the symmetric phase in the form of the current density profiles. In the symmetric phase, the current density reaches its maximum at the wall, but in the broken phase, the maximum takes place at a certain distance from the wall.

Figure~\ref{fig_j_2d_48} also shows that the current maximum occurs at longer distances from the boundary as the magnetic strength increases. However, at the highest magnetic field shown in 
the lowest panel of the same figure, the electric current changes its profile again by shifting its maximum to the boundary in close similarity with the symmetric phase. This sudden change in the behavior is not unexpected since the system experiences the transition to the symmetric phase as the magnetic strength achieves the critical value~\eq{eq_Bc}, and at $B>B_c$, the gauge symmetry is restored. Therefore, the lowest panel of Fig.~\ref{fig_j_2d_48} mimics the results in the symmetric phase qualitatively, Fig.~\ref{fig_j_2d}. Notice that these figures show the normalized electric current, thus highlighting the shape of the current rather than its magnitude.

\begin{figure}[!thb]
\begin{center}
\includegraphics[width=0.5\textwidth,clip=true]{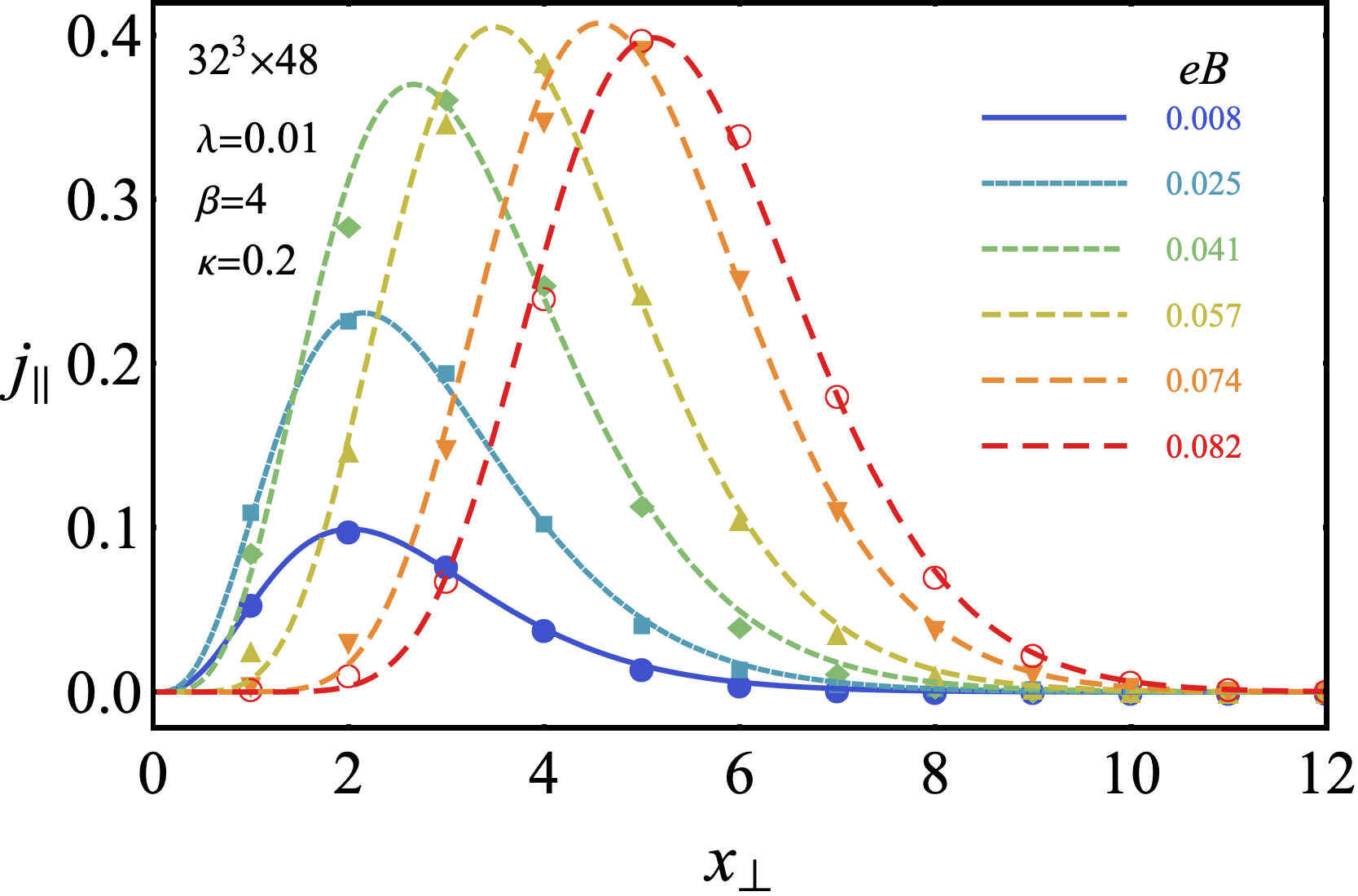}
\end{center}
\vskip -4mm 
\caption{Broken phase~\eq{eq_par_broken}: The electric density current $j_y$ at the function of the distance $x_\perp$ from the boundary at various values of the magnetic field $eB$. The dashed lines correspond to the best fits by function~\eq{eq_j_fit_broken}.}
\label{fig_j_x_48}
\end{figure}

The electric density current $j_y$ as a function of the distance from the boundary is shown in Fig.~\ref{fig_j_x_48} at various magnetic field values below the critical value~\eq{eq_Bc}. This figure shows that the current density profile, which takes its maximum at a certain distance from the wall, can be described very well by the function
\beqn
j^{\mathrm{fit}}_y(x_\perp) =  \frac{M^{1+\nu} |x|^\nu J_{\mathrm{tot}}}{\Gamma(1+\nu) \exp(M x_\perp)} \mathrm{sign}\, x_\perp\,,
\label{eq_j_fit_broken}
\eeqn
where the total current $J_{\mathrm{tot}}$, the effective mass $M$ and the power $\nu$ are the fitting parameters. The prefactors of Eq.~\eq{eq_j_fit_broken} are chosen in such a way that the quantity $J_{\mathrm{tot}}$ corresponds to the total current according to the standard normalization~\eq{eq_int_j}. 

\begin{figure}[!thb]
\begin{center}
\begin{tabular}{cc}
\includegraphics[width=0.45\textwidth,clip=true]{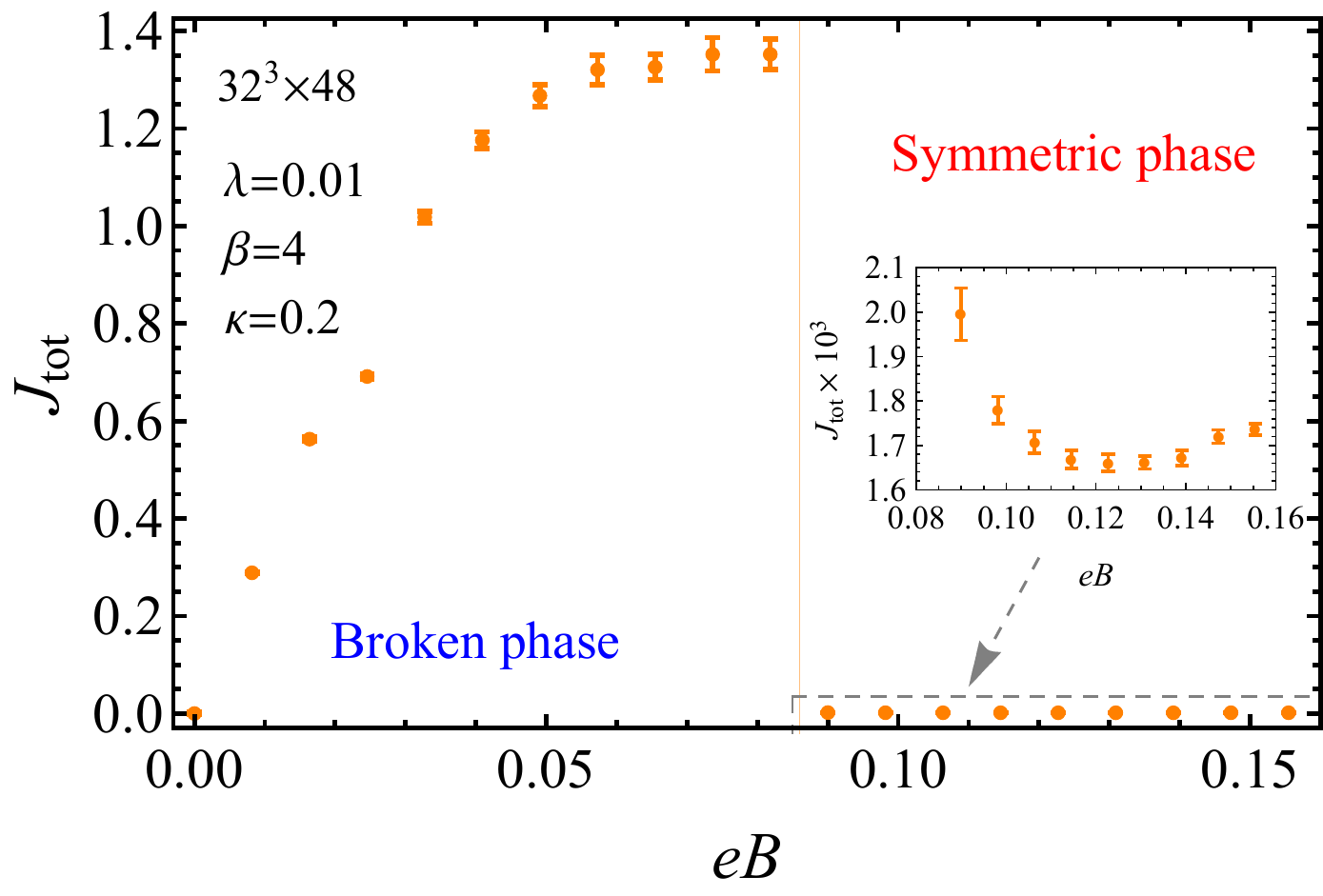} &
\hskip 5mm 
\includegraphics[width=0.45\textwidth,clip=true]{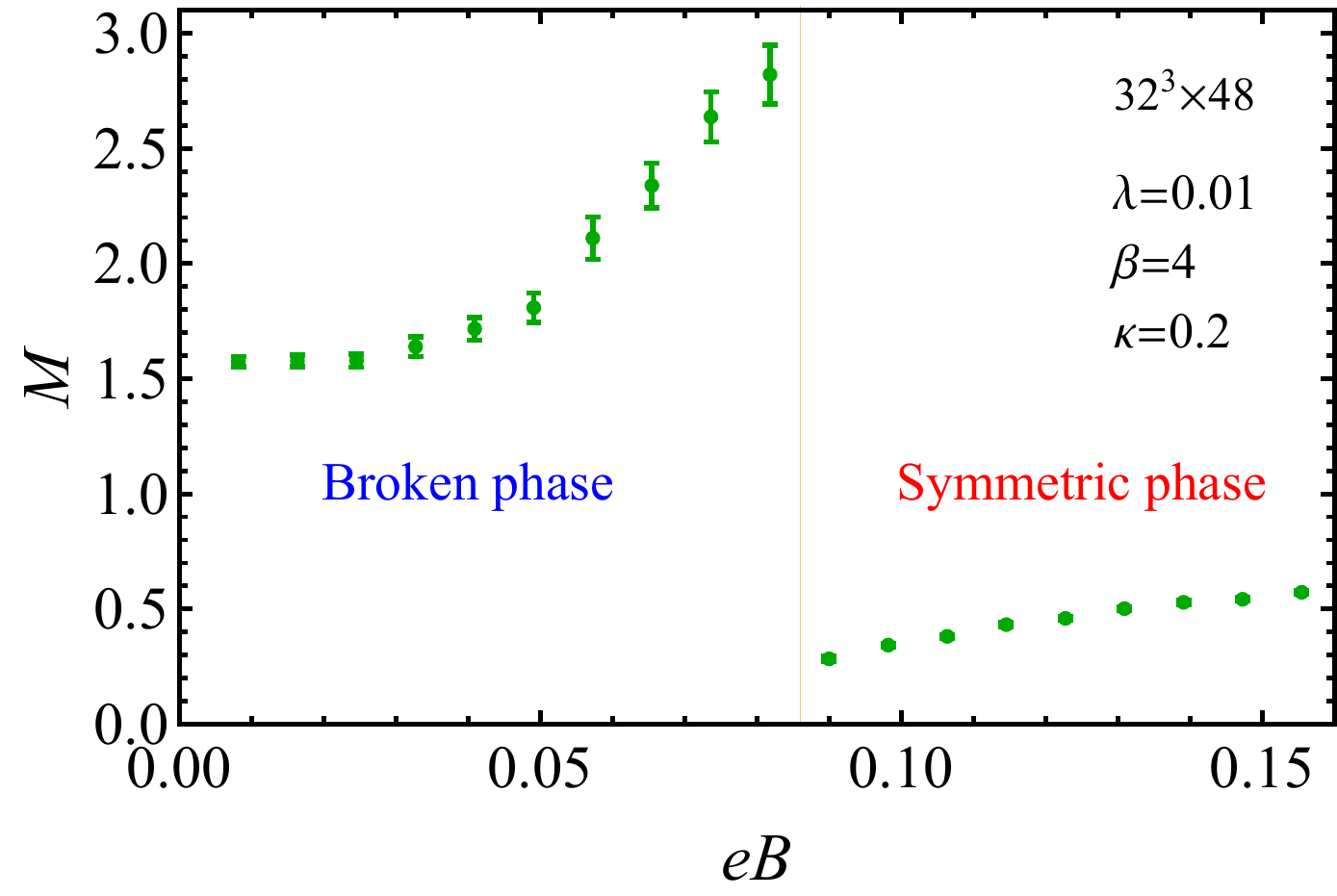} \\
\hskip 10mm (a) & \hskip 15mm (b)
\end{tabular}
\vskip 4mm
\includegraphics[width=0.45\textwidth,clip=true]{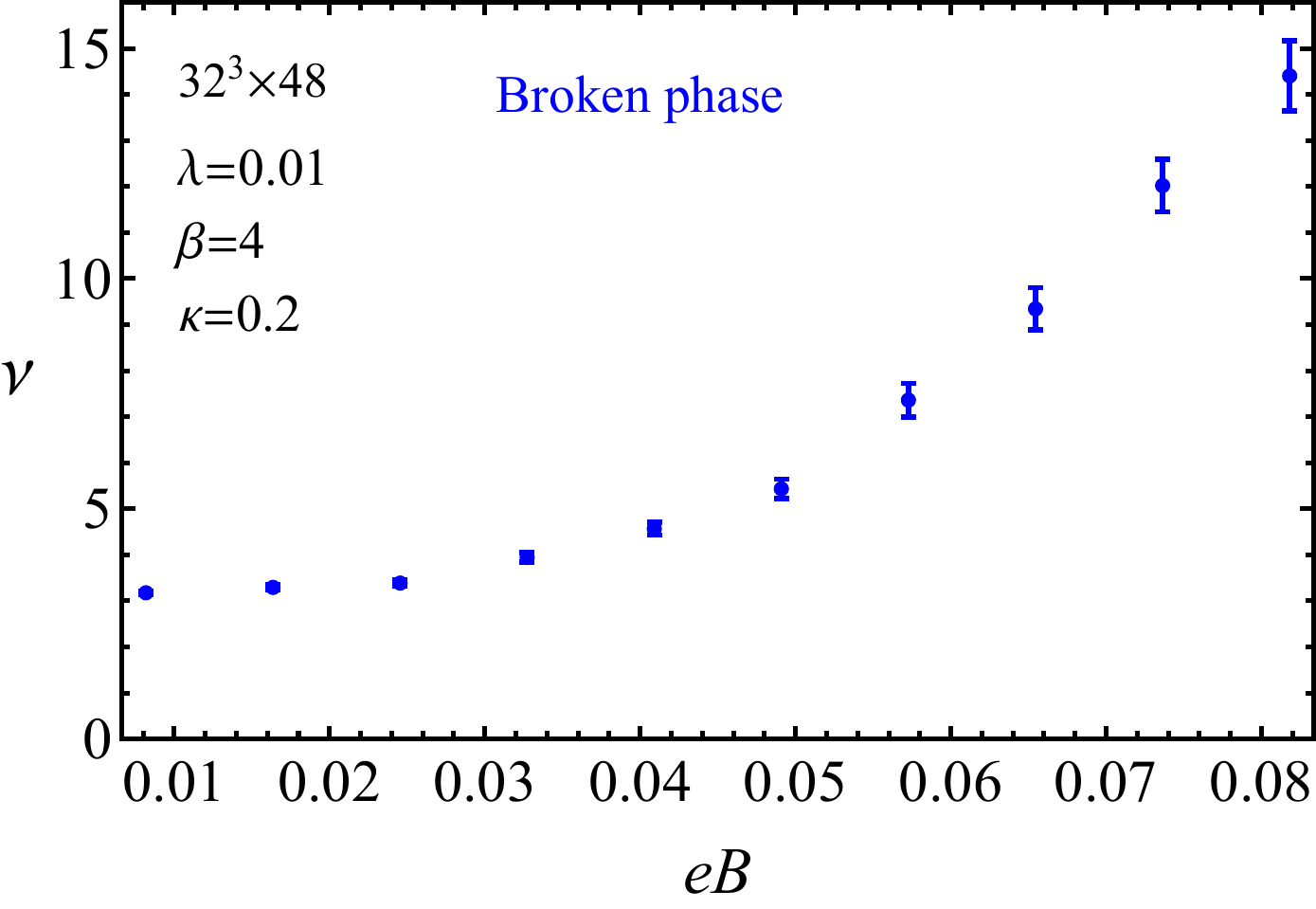} \\
\hskip 7mm (c)
\end{center}
\vskip -4mm 
\caption{Broken phase~\eq{eq_par_broken}: (a) The total current density $J_{\mathrm{tot}}$, (b) effective mass parameter $M$ and (c) the exponent $\nu$ as functions of the magnetic field $eB$. The inset in the upper plot shows the zoom-in on the symmetric phase above the critical magnetic field~\eq{eq_Bc}.}
\label{fig_j_tot_48}
\end{figure}

Figure~\ref {fig_j_x_48} shows that the generated electric current increases in magnitude and takes its maximum further from the wall as the magnetic field increases. This picture works while the strength of the magnetic field still resides below the critical value~\eq{eq_Bc}. The total current, shown in Fig.~\ref{fig_j_tot_48}(a), confirms this observation by showing a linear growth and the saturation of the current below the critical magnetic field~\eq{eq_Bc}, featuring, at the same time, the sudden drop of the current as the magnetic field exceeds the critical value above which the symmetry gets restored. 

Remarkably, Fig.~\ref{fig_j_tot_48} shows that in the symmetry-restored phase, the electric current is three (!) orders of magnitude smaller than in the symmetry-broken phase. In other words, the semi-classical mechanism of the current generation based on the semi-classical vortex solutions drastically exceeds the mechanism based on the quantum vortices, which emerge in the symmetric phase as unstable excitations. Both these mechanisms, illustrated qualitatively in Fig.~\ref{fig_mechanisms}(b), differ from the anomalous current generation catalyzed by the conformal anomaly~\cite{Chu_2018ksb}. 

Notice that the magnitude of the electric current produced in the symmetric phase, which is, in turn, generated by the strong magnetic field -- corresponding to high values of the magnetic field in Fig.~\ref{fig_j_tot_48}(a) -- and the magnitude of the electric current 
in the genuine symmetric phase show in Fig.~\ref{fig_M_J}(b), are the same. This fact highlights some universality of our results, distinguishing clearly the mechanism of the current generation in the symmetric phase from the one in the broken phase. 

Moreover, the total current that emerges in the conformal region -- according to the earlier results of Ref.~\cite{Chernodub_2018ihb} -- appears to be of the same order as the current generated in the symmetric phase. Therefore, while the mechanisms of the current generation in the conformal region and the symmetry restored phase are different -- given by Fig.~\ref{fig_mechanisms}(a) and Fig.~\ref{fig_mechanisms}(b), respectively -- they produce the current of the same magnitude which is, in turn, much smaller than the Meissner current generated in the symmetry broken phase. 

The effective mass obtained from the current distribution with the help of the fit~\eq{eq_j_fit_broken} raises with the increase of magnetic field in agreement with a general tendency~\eq{eq_M_fit} of the mass of the charged scalar particle in the magnetic field background. The mass, shown in Fig.~\ref{fig_j_tot_48}(b), experiences a sudden drop at the critical magnetic field~\eq{eq_Bc} due to the disappearance of the condensate and then again raises with $eB$ following the same qualitative behaviour~\eq{eq_M_fit}.

The power factor $\nu$ of the current density profile~\eq{eq_j_fit_broken} increases with the increase of magnetic field as shown in Fig.~\ref{fig_j_tot_48}(c), raising from a modest $\nu \simeq 3$ to a very high value $\nu \simeq 15$ close to the transition point~\eq{eq_Bc}.

\begin{figure}[!thb]
\begin{center}
\begin{tabular}{cc}
\includegraphics[width=0.45\textwidth,clip=true]{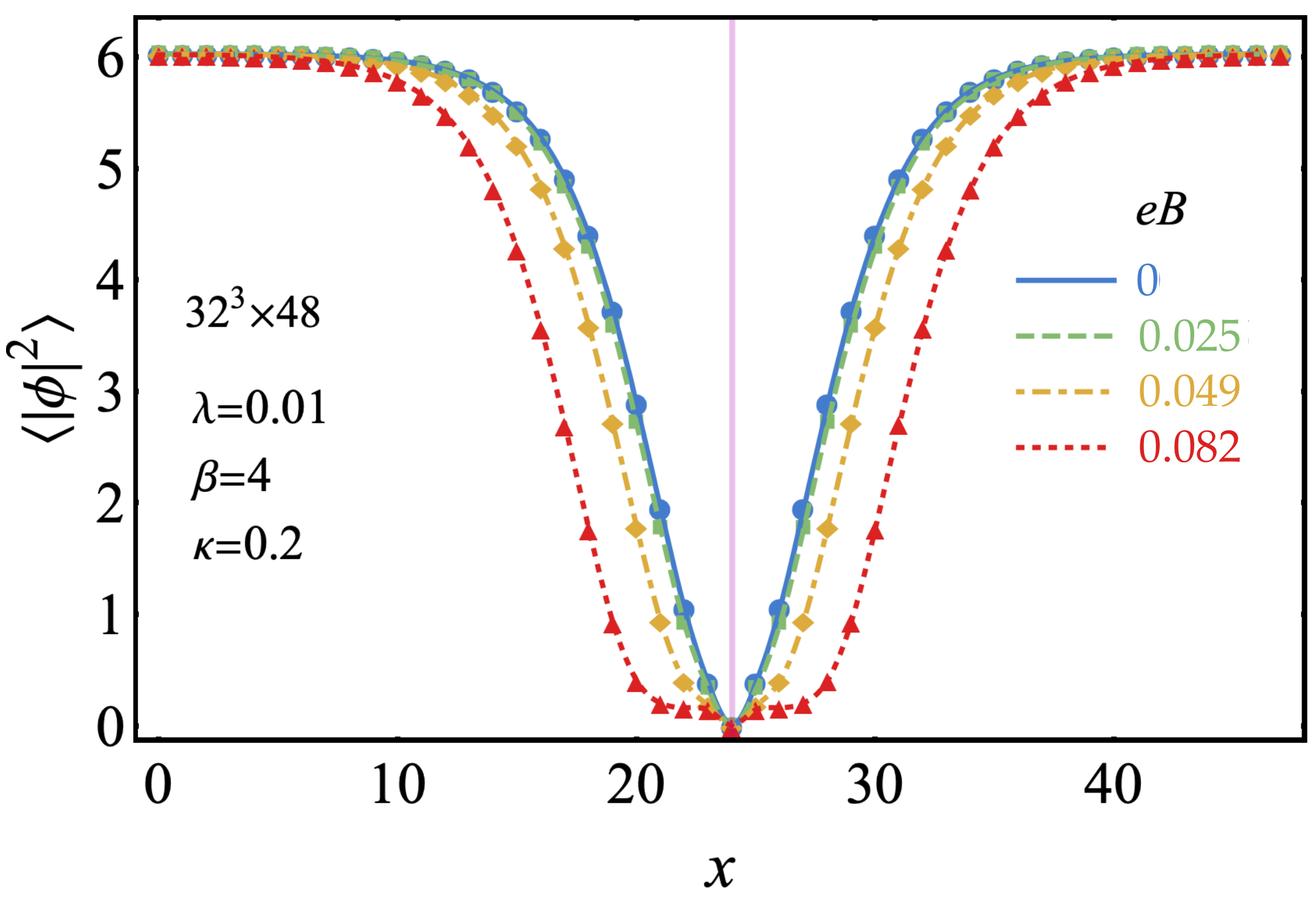} &
\hskip 5mm 
\includegraphics[width=0.45\textwidth,clip=true]{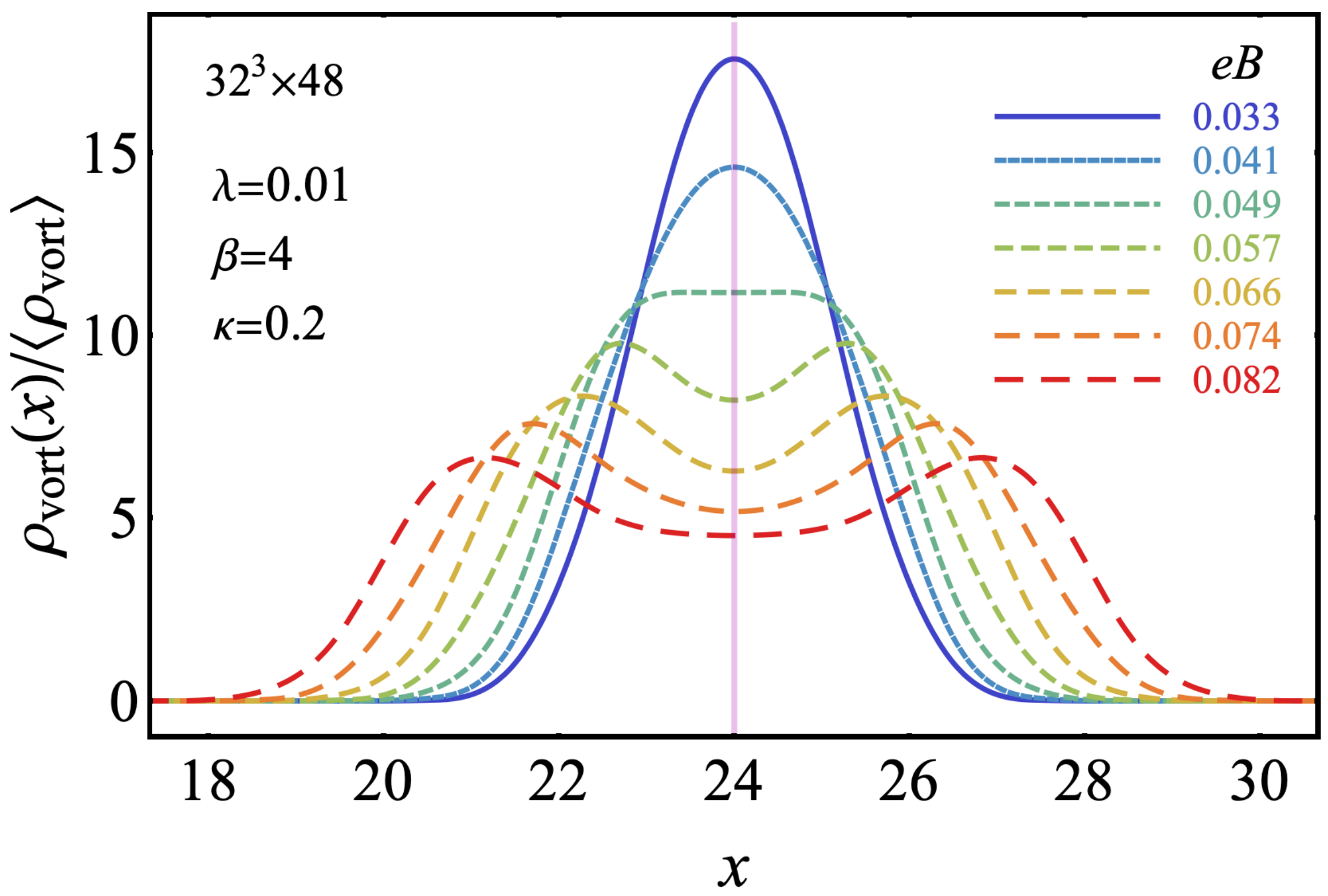} \\
\hskip 10mm (a) & \hskip 15mm (b)
\end{tabular}
\end{center}
\vskip -4mm 
\caption{Broken phase~\eq{eq_par_broken}: (a) the scalar field fluctuations $\avr{|\phi|^2}$ and (b) the normalized vortex density at a few values of the background magnetic field.}
\label{fig_vort_48}
\end{figure}

Finally, in Figs.~\ref{fig_vort_48}(a) and (b), we show the behavior of the scalar field and the normalized vortex density, respectively, as functions of the normal distance to the wall. The increasing magnetic field widens the hollow in the condensate near the boundary and creates a double layer of vortices near the wall. This nontrivial structure appears due to the specific geometry of our problem and the mutual repulsion of the vortices interacting through the wall. In addition, the vortices create a strong electric Meissner current due to the mechanism depicted in Fig.~\ref{fig_mechanisms}(b).

\section{Summary and Conclusions}

Our paper is concentrated on first-principle numerical results for electric current induced by a background magnetic field close to a boundary in the vacuum of the Abelian Higgs model. This model represents a scalar counterpart of quantum electrodynamics with charged scalar fields instead of electrically charged spinors (therefore, it is also called ``scalar QED''). Our investigation is carried out at zero temperature.

In the conformal limit of the zero-temperature model, the boundary current arises from quantum fluctuations~\cite{McAvity:1990we,Chu_2018ksb,Chu_2018ntx} due to a simple mechanism~\cite{Chu_2018ksb}. Initially, vacuum fluctuations generate particle-anti-particle pairs which move along closed circular paths in a magnetic field background and then annihilate. However, their mutual orbit remains open when a virtual particle-anti-particle pair is formed near a reflective boundary with the magnetic field oriented tangentially to the boundary. Consequently, the virtual particles follow skipping orbits without fast annihilation, thus establishing the generation of electric current along the boundary. The masslessness of particles facilitates the creation of particle pairs from the vacuum, amplifying the boundary current and enforcing its long-range nature. The strength of the current is proportional to the beta function responsible for the renormalization of the electric charge.

Outside the conformal limit, the scalar electrodynamics can reside either in the symmetry-restored phase featuring a massive scalar field or in the regime with spontaneous symmetry breaking in which the charged scalar field is condensed. We found that in the symmetry-unbroken regime, the current-generation effect gets suppressed in its amplitude far from the boundary because of the short-range nature of fluctuations of the massive virtual particles in agreement with the analytical expectations~\cite{Hu:2020puq}. However, the total current that emerges in the conformal region appears to be of the same order as the current generated in the symmetric phase.

Our numerical calculations show that the quantum boundary current gets drastically enhanced in the symmetry-broken phase due to the condensation of charged particles. We consider the reflective boundary condition of the Dirichlet type, which enforces the vanishing of the scalar field at the boundary and is qualitatively different from the superconducting (Neumann-type) boundary conditions. 

We consider a superconducting regime with a transparent semi-classical vortex-based mechanism to generate the non-conformal regime's boundary current. We find that the vortices are attracted to the boundary, and their circular currents at up coherently, thus generating the global magnetization current along the boundary of the material. The mechanism is similar to the standard Meissner effect in which the vortices are forced to flow around the surface, creating a circulating current that shields the superconductor from the external magnetic field~\cite{Tinkham2004}. As the superconducting condensate is forced to vanish at the boundary, the vortices near the boundary are more energetically favorable than in the bulk of the material.  

The boundary current in the conformal regime possesses a much smaller amplitude than the current in the symmetry-broken phase since the former effect is a purely quantum phenomenon while the latter has a semi-classical nature. Despite the fact that the origin and magnitude of these effects differ, they lead to the same phenomenon, the dissipationless magnetization current along the boundary. 

Finally, we would like to discuss experimentally observable consequences of the induced magnetization current due to the conformal anomaly in Dirac semimetals. In one-component massless QED, the magnetization current~\eq{eq_j_conformal} leads to the enhanced edge conductivity which can be estimated as
\begin{align}
    \delta \sigma_{xy} = \frac{2 \beta}{e \hbar} \ln \frac{L_{\mathrm{IR}}}{L_{\mathrm{UV}}}\,,
    \label{eq_delta_sigma}
\end{align}
where $L_{\mathrm{IR}}$ and $L_{\mathrm{UV}}$ are infrared and ultraviolet distance scales, respectively. Outside the conformal limit, which is typical for most of the Dirac semimetals, the former length scale is inversely proportional to the mass which breaks the scale invariance. The ultraviolet length scale in Eq.~\eq{eq_delta_sigma} can be taken to be the inter-atomic distance~$L_{\mathrm{UV}} \simeq a$, which can be taken a few Angstroms. Taking, for the sake of estimation, the mass scale to be equal about $10$\,K (in temperature units) and $a$ of the order of a few Angstrom, one can estimate the logarithmic factor in Eq.~\eq{eq_delta_sigma} as $\ln (L_{\mathrm{IR}}/L_{\mathrm{UV}}) \simeq 4\pi$, which gives us 
\begin{align}
    \delta \sigma_{xy} = \frac{4 c}{3 \varepsilon v_F} \sigma^{\mathrm{Hall}}\,,
    \label{eq_delta_sigma_estimation}
\end{align}
where $\sigma^{\mathrm{Hall}} = e^2/h$ is the Hall conductivity. Due to the nature of logarithm function, a order-of-magnitude change in $L_{\mathrm{IR}}$ does not affect the prefactor of Eq.~\eq{eq_delta_sigma_estimation} significantly. Notice that the QED beta function is four times bigger than the one in the scalar QED~\eq{eq_beta_function}.

In Eq.~\eq{eq_delta_sigma_estimation}, which takes peculiarities of the running of electric charge in Dirac semimetals~\cite{Isobe2012}, $\varepsilon \simeq 10$ is a dielectric constant and $v_F \simeq c/300$ is the Fermi velocity. The slowness of electrons enhances the beta function and makes significant contribution to the magnetization current resulting in the large value of conductivity associated with the magnetization current:
\begin{align}
    \delta \sigma_{xy} \simeq 40 \sigma^{\mathrm{Hall}}\,.
\label{eq_estimation}
\end{align}
Notice that this estimation~\eq{eq_estimation} does not refer to the Hall effect but to the magnetization current, which is confined to the boundaries to material and, therefore, does not generate any net transverse current.

The persistent boundary current creates accompanying magnetic field which can be probed, for example, using scanning probe based on a nitrogen-vacancy (NV) defect in a diamond as an atomic-size quantum magnetometer~\cite{Maletinsky2012}. The NV-based measurement is very sensitive to the magnetic field through the Zeeman effect, which allowed, for example, to probe the hydrodynamic transport of electrons in bulk crystals WTe${}_2$~\cite{Vool2021} at low temperatures. Another, perhaps, less complicated option is to measure the enhanced local conductivity at the edge of the crystal using microwave impedance microscopy which has been employed to observe the quantum spin Hall edges in monolayer WTe${}_2$ in Ref.~\cite{Shi2019}. 

\medskip
\textbf{Acknowledgements} 
\par
The work of VAG and AVM was supported by Grant No. 0657-2020-0015 of the Ministry of Science and Higher Education of Russia. The numerical simulations were performed at the computing cluster Vostok-1 of Far Eastern Federal University.

\medskip


\begin{thebibliography}{10}
\providecommand{\url}[1]{\texttt{#1}}
\providecommand{\urlprefix}{URL }

\bibitem{Kharzeev:2012ph}
D.~E. Kharzeev, K.~Landsteiner, A.~Schmitt, H.-U. Yee,
\newblock \emph{Lect. Notes Phys.} \textbf{2013}, \emph{871} 1.

\bibitem{Kharzeev:2013ffa}
D.~E. Kharzeev,
\newblock \emph{Prog. Part. Nucl. Phys.} \textbf{2014}, \emph{75} 133.

\bibitem{Landsteiner:2011cp}
K.~Landsteiner, E.~Megias, F.~Pena-Benitez,
\newblock \emph{Phys. Rev. Lett.} \textbf{2011}, \emph{107} 021601.

\bibitem{Landsteiner:2012kd}
K.~Landsteiner, E.~Megias, F.~Pena-Benitez,
\newblock \emph{Lect. Notes Phys.} \textbf{2013}, \emph{871} 433.

\bibitem{Huang:2015oca}
X.-G. Huang,
\newblock \emph{Rept. Prog. Phys.} \textbf{2016}, \emph{79}, 7 076302.

\bibitem{Landsteiner:2016led}
K.~Landsteiner,
\newblock \emph{Acta Phys. Polon. B} \textbf{2016}, \emph{47} 2617.

\bibitem{Capper:1973mv}
D.~M. Capper, M.~J. Duff,
\newblock \emph{Nucl. Phys. B} \textbf{1974}, \emph{82} 147.

\bibitem{Capper:1974ed}
D.~M. Capper, M.~J. Duff, L.~Halpern,
\newblock \emph{Phys. Rev. D} \textbf{1974}, \emph{10} 461.

\bibitem{Deser:1976yx}
S.~Deser, M.~J. Duff, C.~J. Isham,
\newblock \emph{Nucl. Phys. B} \textbf{1976}, \emph{111} 45.

\bibitem{Chernodub:2021nff}
M.~N. Chernodub, Y.~Ferreiros, A.~G. Grushin, K.~Landsteiner, M.~A.~H.
  Vozmediano,
\newblock \emph{Phys. Rept.} \textbf{2022}, \emph{977} 1.

\bibitem{Chernodub:2016lbo}
M.~N. Chernodub,
\newblock \emph{Phys. Rev. Lett.} \textbf{2016}, \emph{117}, 14 141601.

\bibitem{Luttinger:1964zz}
J.~M. Luttinger,
\newblock \emph{Phys. Rev.} \textbf{1964}, \emph{135} A1505.

\bibitem{Chernodub:2017jcp}
M.~N. Chernodub, A.~Cortijo, M.~A.~H. Vozmediano,
\newblock \emph{Phys. Rev. Lett.} \textbf{2018}, \emph{120}, 20 206601.

\bibitem{Chernodub:2019tsx}
M.~N. Chernodub, C.~Corian\`o, M.~M. Maglio,
\newblock \emph{Phys. Lett. B} \textbf{2020}, \emph{802} 135236.

\bibitem{Bermond:2022mjo}
B.~Bermond, M.~Chernodub, A.~G. Grushin, D.~Carpentier \textbf{2022}.

\bibitem{Northe:2022tjr}
C.~Northe, C.~Zhang, R.~Wawrzy\'nczak, J.~Gooth, S.~Galeski, E.~M. Hankiewicz
  \textbf{2022}.

\bibitem{Chu:2018ksb}
C.-S. Chu, R.-X. Miao,
\newblock \emph{Phys. Rev. Lett.} \textbf{2018}, \emph{121}, 25 251602.

\bibitem{Chu:2018ntx}
C.-S. Chu, R.-X. Miao,
\newblock \emph{JHEP} \textbf{2018}, \emph{07} 005.

\bibitem{McAvity:1990we}
D.~M. McAvity, H.~Osborn,
\newblock \emph{Class. Quant. Grav.} \textbf{1991}, \emph{8} 603.

\bibitem{Chernodub:2019blw}
M.~N. Chernodub, M.~A.~H. Vozmediano,
\newblock \emph{Phys. Rev. Research.} \textbf{2019}, \emph{1} 032002.

\bibitem{Herzog:2017xha}
C.~P. Herzog, K.-W. Huang,
\newblock \emph{JHEP} \textbf{2017}, \emph{10} 189.

\bibitem{Andrei:2018die}
N.~Andrei, et~al.,
\newblock \emph{J. Phys. A} \textbf{2020}, \emph{53}, 45 453002.

\bibitem{Chu:2020mwx}
C.-S. Chu, R.-X. Miao,
\newblock \emph{Phys. Rev. D} \textbf{2020}, \emph{102}, 4 046011.

\bibitem{Hu:2020puq}
P.-J. Hu, Q.-L. Hu, R.-X. Miao,
\newblock \emph{Phys. Rev. D} \textbf{2020}, \emph{101}, 12 125010.

\bibitem{Miao:2022oas}
R.-X. Miao, Y.-Q. Zeng,
\newblock \emph{Phys. Lett. B} \textbf{2023}, \emph{838} 137700.

\bibitem{Tinkham2004}
M.~Tinkham,
\newblock \emph{Introduction to Superconductivity},
\newblock Dover Books on Physics Series. Dover Publications, \textbf{2004}.

\bibitem{Chernodub_2018ihb}
M.~N. Chernodub, V.~A. Goy, A.~V. Molochkov,
\newblock \emph{Phys. Lett. B} \textbf{2019}, \emph{789} 556.

\bibitem{Chu_2018ksb}
C.-S. Chu, R.-X. Miao,
\newblock \emph{Phys. Rev. Lett.} \textbf{2018}, \emph{121}, 25 251602.

\bibitem{Nakayama:2013is}
Y.~Nakayama,
\newblock \emph{Phys. Rept.} \textbf{2015}, \emph{569} 1.

\bibitem{Boyd:1996bx}
G.~Boyd, J.~Engels, F.~Karsch, E.~Laermann, C.~Legeland, M.~Lutgemeier,
  B.~Petersson,
\newblock \emph{Nucl. Phys. B} \textbf{1996}, \emph{469} 419.

\bibitem{Ranft:1982hf}
J.~Ranft, J.~Kripfganz, G.~Ranft,
\newblock \emph{Phys. Rev. D} \textbf{1983}, \emph{28} 360.

\bibitem{Gattringer2009}
C.~Gattringer, C.~Lang,
\newblock \emph{Quantum chromodynamics on the lattice: an introductory
  presentation}, volume 788,
\newblock Springer Science \& Business Media, \textbf{2009}.

\bibitem{Omelyan2002}
I.~Omelyan, I.~Mryglod, R.~Folk,
\newblock \emph{Physical Review E} \textbf{2002}, \emph{65}, 5 056706.

\bibitem{Bali:2011qj}
G.~S. Bali, F.~Bruckmann, G.~Endrodi, Z.~Fodor, S.~D. Katz, S.~Krieg,
  A.~Schafer, K.~K. Szabo,
\newblock \emph{JHEP} \textbf{2012}, \emph{02} 044.

\bibitem{Einhorn:1977qv}
M.~B. Einhorn, R.~Savit,
\newblock \emph{Phys. Rev. D} \textbf{1978}, \emph{17} 2583.

\bibitem{Chu_2018ntx}
C.-S. Chu, R.-X. Miao,
\newblock \emph{JHEP} \textbf{2018}, \emph{07} 005.

\bibitem{Isobe2012}
H.~Isobe, N.~Nagaosa,
\newblock \emph{Physical Review B} \textbf{2012}, \emph{86}, 16.

\bibitem{Maletinsky2012}
P.~Maletinsky, S.~Hong, M.~S. Grinolds, B.~Hausmann, M.~D. Lukin, R.~L.
  Walsworth, M.~Loncar, A.~Yacoby,
\newblock \emph{Nature Nanotechnology} \textbf{2012}, \emph{7}, 5 320.

\bibitem{Vool2021}
U.~Vool, A.~Hamo, G.~Varnavides, Y.~Wang, T.~X. Zhou, N.~Kumar, Y.~Dovzhenko,
  Z.~Qiu, C.~A.~C. Garcia, A.~T. Pierce, J.~Gooth, P.~Anikeeva, C.~Felser,
  P.~Narang, A.~Yacoby,
\newblock \emph{Nature Physics} \textbf{2021}, \emph{17}, 11 1216.

\bibitem{Shi2019}
Y.~Shi, J.~Kahn, B.~Niu, Z.~Fei, B.~Sun, X.~Cai, B.~A. Francisco, D.~Wu, Z.-X.
  Shen, X.~Xu, D.~H. Cobden, Y.-T. Cui,
\newblock \emph{Science Advances} \textbf{2019}, \emph{5}, 2.

\end{thebibliography}

\end{document}